\documentstyle[12pt,preprint,eqsecnum,aps,epsfig,floats]{revtex}
\begin{document}
\input epsf

\tighten
\draft
\preprint{}
\def\eg{{\it e.g.}}
\def\today{\number\month/\number\day/\number\year}

\title{Classical Dynamics of the \\
 Quantum Harmonic 
Chain}
\author{Todd Brun\thanks{Current address:  Department of Physics,
Carnegie Mellon University, Pittsburgh, PA 15213.
Email:  tbrun@andrew.cmu.edu}}

\address{\setlength{\baselineskip}{.2in} \sl Institute of Theoretical
Physics\\
University of California,\\
Santa Barbara, CA 93106-4030}

\author{James B.~Hartle\thanks{hartle@cosmic.physics.ucsb.edu}}

\address{\setlength{\baselineskip}{.2in}
\sl Department of Physics,\\
University of California,\\
Santa Barbara, CA 93106-9530}

\maketitle

\begin{abstract}

The origin of classical predictability is investigated
for the one dimensional harmonic chain considered as a closed
quantum mechanical system.  By comparing the properties
of a family of coarse-grained descriptions of the chain, 
we conclude that local coarse-grainings in this family are more useful
for prediction than nonlocal ones.   A quantum mechanical system 
exhibits classical behavior when
the probability is high for histories
having the correlations in time implied by classical deterministic
laws. But approximate classical determinism holds only for  certain 
coarse-grainings and then only if the initial state of the system is
suitably restricted. Coarse-grainings by the values of the 
hydrodynamic variables (integrals over suitable
volumes of densities of approximately conserved quantities) 
define the histories  usually used in classical physics. But what
distinguishes this coarse-graining from others? This paper
approaches this question by analyzing a family of coarse-grainings
for the linear harmonic chain.  At one extreme in the family 
the chain is divided into local groups of $N$ atoms. At the other
extreme the $N$ atoms are distributed nonlocally over the whole
chain. Each coarse-graining follows the average (center of mass) positions 
of the groups and ignores the ``internal'' 
coordinates within each group, these constituting  a different
environment for each coarse-graining.
For an initial condition where long wavelength modes
are excited and short wavelength modes are distributed thermally 
we find that the coarse-grained positions obey deterministic
equations of motion accompanied by noise. The noise is greater
the more nonlocal the coarse-graining. Further, the deterministic
equations require more time steps to evolve over a given time interval
for the nonlocal coarse-grainings than for the local ones. A continuum
limit is possible only for the near local coarse-grainings.  For 
parameters of the model characteristic of realistic situations these
features strongly favor the local coarse-grainings over the nonlocal
ones for prediction.
Each of these differences can be traced to the approximate conservation
of the local center of mass momentum. 
We then consider the chain quantum mechanically
and show that for realistic parameters, all the coarse-grainings decohere
rapidly compared to dynamical time scales.
We conclude that noise, decoherence, and computational complexity
favor locality over nonlocality for deterministic predictability.

\end{abstract}

\date{\today}

\pacs{}

\tighten

\setcounter{footnote}{0}
\section{Introduction}
\label{sec: I}

As far as we know them, the laws of physics that apply universally
to all physical systems are quantum mechanical. 
The universe  at a fundamental level is therefore characterized by
indeterminacy and distributed probabilities.  The wide
range of applicability of classical deterministic laws
is an empirical fact to be explained from the
universe's quantum dynamics and initial quantum state. This paper investigates the
origin of classical predictability for the very simple model of a
linearized chain of idealized atoms in the context of the quantum mechanics
of closed systems \cite{Grif,Omn94,GH90a}, most generally 
quantum cosmology.
We exhibit decoherent sets of 
coarse-grained histories for
which, given suitable restrictions on the state, the probability is high
for histories exhibiting
the correlations in time governed by the classical wave equation. We
shall compare these quasiclassical coarse-grainings with a class of others
and analyze why the classical coarse-grainings are the most predictable in
the class.

Why do we raise the question of the origin of classical predictability over
seventy years after the initial formulation of quantum mechanics? Every quantum
mechanics text contains some treatment of this question. Ehrenfest's
theorem is the starting point for one such discussion.  For a
nonrelativistic particle of mass $m$ moving in one dimension in a
potential $V(x)$,  Ehrenfest's theorem is the exact relation between expected values:
\begin{equation}
m\ \frac{d^2\langle x\rangle}{dt^2} = - \left\langle\frac{\partial
V(x)}{\partial x}\right\rangle\ .
\label{oneone}
\end{equation}
This is not a deterministic equation of motion, but 
for certain states, typically narrow wave packets, the expected value of
the force may approximated by the force at the expected value of the
position, thereby giving a classical equation of motion for that expected
value:
\begin{equation}
m\ \frac{d^2\langle x\rangle}{dt^2} = - \frac{\partial
V(\langle x\rangle)}{\partial x}\ .
\label{onetwo}
\end{equation}
This equation shows that the orbit of the center of a narrow wave packet
obeys Newton's laws. 

This kind of elementary derivation already exhibits two necessary
requirements for a quantum system to exhibit classical deterministic
behavior. Some coarseness in the description is needed, as well as a
restriction on the initial quantum state. However, otherwise this kind of
demonstration does not address the issues we hope to discuss in quantum
cosmology for the following reasons:
\begin{itemize}
\item The behavior of expected values in time is not enough to define
classical behavior. Equations of motion predict {\it correlations}
in time, which
in quantum mechanics are properties of the probabilities for time {\it
histories}. The statement that the earth moves on a classical orbit is
most correctly understood in quantum theory as the assertion that, among a
decoherent set of coarse-grained histories of the earth's position in time,
the probability is high for histories exhibiting the deterministic
correlations in time implied by Newton's laws and low for all others.
To discuss classical predictability therefore we should be dealing with the
probabilities of time histories, not merely with the time dependence of
expected values.

\item The Ehrenfest derivation relies on a close connection between the
equations of motion arising from the
fundamental action and the phenomenological
equations of motion determining classical correlations. There is no such
connection in general. In general situations we expect classical equations
of motion like the Navier-Stokes equation relating values of continuum
hydrodynamic variables at different times, incorporating phenomenological
equations of state, and exhibiting dissipation, noise, and irreversibility.
The equations of the fundamental theory, whether one takes it to be quantum
electrodynamics or $M$-theory, exhibit none of these phenomena and are at
best only distantly related in form. We need to derive the {\it form} of
the classical equations of motion as well as the probabilities with which
they are satisfied.

\item The Ehrenfest derivation posits the variable---the position
$x$---in which classical deterministic behavior is exhibited. But the quantum
mechanics of any closed system will exhibit a great many 
complementary sets of decoherent histories some of which may exhibit
deterministic correlations in time. What distinguishes coarse-grainings in
terms of the familiar quasiclassical variables from all other possibilities
exhibited by a closed quantum mechanical system? Certainly it is not their
relation to the variables of the fundamental theory, which is  
typically only distant, as described above.
Rather, it must lie in the relative utility of
different coarse-grainings for prediction, with quasiclassical
variables being highly predictable.  A complete derivation of
classical predictability must seek to distinguish classical
coarse-grainings from all others.

\item The Ehrenfest derivation deals with the
expected outcomes of ``measurements'' on
an otherwise isolated subsystem. However, in quantum cosmology we are
interested in classical behavior in much more general situations, over
cosmological stretches of space and time and over a wide range of systems
including the universe as a whole, whether or not they are receiving the
attention of observers. We are interested in 
sets of alternative histories that can be assigned probabilities whether or not
they describe measurement situations; in the quantum mechanics of
closed systems that means the sets must decohere.
Decoherence is thus a prerequisite for classical behavior.
\end{itemize}

Histories of the quasiclassical domain of everyday experience are
coarse-grained by values of usual quasiclassical variables.  These
include various sorts of hydrodynamic
variables---averages over suitably small volumes of
densities of conserved or nearly conserved quantities. Densities of energy,
momentum, baryon number, nuclear and chemical species are examples. The
system behaves classically when the probability is high for histories that
exhibit correlations in time summarized by phenomenological classical
equations of motion such as the Navier-Stokes equation. 

Simple arguments \cite{GH90a,GH93a,Bru93,BH96a,Hal98,JJH99a}
suggest why histories of
these  hydrodynamic, quasiclassical variables should decohere and exhibit
classical correlations in time. Coarse-graining is generally necessary for
decoherence. Roughly speaking, a coarse-graining  divides 
the variables of the system into
those that are followed by  the coarse-graining and those that are ignored. 
The ignored variables constitute the environment.
Interaction between these classes of variables
is necessary to dissipate the phases
between different coarse-grained histories and to achieve decoherence.
However, that same interaction produces noise, which causes deviations from
predictability. Integrals of densities of conserved or approximately
conserved quantities are natural candidates for quasiclassical variables.
Their approximate conservation enables them to resist deviations from
predictability caused by the noise arising from their interactions with the
rest of the universe. Further, following standard arguments of
nonequilibrium statistical mechanics, their
approximate conservation  leads to correlations in time summarized by a
closed set of equations of motion. All isolated systems approach
equilibrium. However, averages of approximately conserved quantities
over suitable volumes approach equilibrium slowly. Closed sets
of equations of motion result when the volumes can be chosen large enough
that statistical fluctuations and noise are unimportant, but small enough
that equilibrium is established within each volume in a time short compared
to the dynamical time scales on which the variables vary (see, \eg
\cite{Zub74}). The
constitutive relations defining equations of state, coefficients of
viscosity, diffusion, {\it etc}., are then defined, permitting closure
of a set of hydrodynamic equations of motion. Local
equilibrium being thus established, the further equilibration of the
volumes among themselves is governed by these equations.

Despite the plausibility of the above general and simple qualitative
picture, its validity has been only partially investigated quantitatively. 
To make the argument quantitative, in light of our earlier discussion,
requires at a minimum an investigation of 
sets of histories coarse-grained by ranges of quasiclassical variables
which has the following features:

\begin{itemize}
\item Establishes the decoherence of sets of histories sufficiently
      coarse-grained by ranges of quasiclassical variables.

\item Establishes with high probability deterministic correlations in 
      time summarized by closed systems of classical equations of motion
      for reasonably realistic initial conditions. 

\item Compares the decoherence and predictability of different
      coarse-grained sets of histories, both within the family of
      coarse-grainings by quasiclassical variables and with other
      coarse-grainings of distinct character.

\end{itemize}

Despite intensive investigation of all of these points separately there
is as yet no analysis which combines all three.
There are
many investigations of the mechanisms of decoherence of histories 
\cite{DH92,GH93a,Bru93,HZ97} and
of the closely related decoherence of density matrices \cite{decdm}, especially in linear
systems.  However, these studies have typically posited a fixed 
division of the
fundamental variables into those describing a ``system'' and those
describing its ``environment''. Coarse-grainings follow variables of the
system while ignoring those constituting the environment. Such a
fixed system-environment split is intuitively accessible and correctly
models many mechanisms of decoherence, but it is not general. 

Rather, coarse-graining is the general notion which, when possible, 
determines a family of system-environment splits. 
Different coarse-grainings
lead to different possible notions of system and environment, that
division is not usually unique, and for some kinds of coarse-graining no
system-environment split is possible at all. (See Appendix A.)  
Even when a system-environment split is possible at one time, {\it
different } system-environment splits could be needed at other 
moments of time. A fixed system-environment split is therefore neither
general nor necessary for formulation quantum mechanics.  However, as
such workers as Feynman and Vernon \cite{FV63},
Joos and Zeh \cite{JZ85}, Zurek \cite{Zursum},
Caldeira and Leggett \cite{CL83}, and Omn\`es \cite{Omn97} fully
appreciated, a system-environment split is an important tool for analyzing
specific coarse-grainings and for understanding the physical mechanisms of
decoherence. We shall utilize this tool extensively is what follows.

The emergence of deterministic correlations in time governed by
classical equations of motion has been investigated for fixed
system-environment splits \cite{GH93a}. In a recent elegant paper,
Halliwell \cite{Hal98} has derived classical equations of motion for 
hydrodynamic variables although with a special assumption about 
the nature of the environment. Neither of these works compared
quasiclassical coarse-grainings with nonclassical ones.
In their pioneering paper on classical behavior in systems of 
interacting spins, Brun and Halliwell \cite{BH96a} compared a family of
coarse-grainings but did not derive classical equations of motion.

This paper moves the analysis of decoherence and classicality a step
towards realistic coarse-grainings by hydrodynamic variables
in the context of a
simple model---a one-dimensional chain of point masses with linear,
nearest-neighbor interactions. (See Figure 1).
This is the simplest model which exhibits
time correlations governed by a continuum equation of motion---the wave
equation---in appropriate coarse-grainings for certain initial states.
Yet the Lagrangian of the model is quadratic in all coordinates, so it
can be tractably analyzed with standard path integral techniques
\cite{FV63,CL83,GH93a}. Although our analysis is only for this 
very simple system, and does not deal with coarse-grainings by
hydrodynamic variables {\it per se}, it does display all three
features listed above. 

The details of the model are laid out in Section
II, where we introduce a family of coarse-grainings.  In the simplest,
the chain of ${\cal N}$ identical particles is divided into ${\cal M}$ groups,
each consisting of $N$ neighboring atoms.  Histories are partitioned by
ranges of values of the center of mass displacement in each group.
We find that the effective equations of motion for these variables
are well approximated by the classical wave equation.  We obtain
quantitative estimates for the statistical noise causing deviations from
the predictions of the wave equations. 

We compare this quasiclassical coarse-graining with other members of
a family of sets of alternative coarse-grained histories constructed
as shown in Figure 2. The chain is divided up into $\cal M$
groups of $N$ atoms, each group consisting of equally spaced clumps of $d$ 
atoms. 
We coarse-grain by equal ranges $\Delta$ of the values of the
{\it average (center of mass) positions of the atoms in a group} at 
equally spaced intervals of time $\Delta t$. A family
of sets of alternative coarse-grained  histories is thereby defined
parameterized by $(N, d, \Delta, \Delta t)$. When $d=N$ the $N$ atoms
in a group are all neighbors, the coarse-graining entirely local,
and the average position related to the approximately conserved
center of mass momentum of the group. There is nothing special about
the remaining members of the family except that they are amenable to
analysis and range from local to highly nonlocal as $d$ decreases
from $N$ to $1$. 

We find that for each member of this family that decoheres the 
probabilities of the histories can be thought of as obeying
classical equations of motion augmented by noise [{\it cf}
(\ref{threeten})]. 
We compare the members of this family with respect to three properties
bearing on classical behavior: decoherence, noise,
and the computational complexity of their equations of motion.
We find marked differences which are describe in detail in
the Conclusion, but which we summarize briefly here. As measured
by the smallness of the ratio of the decoherence to dynamical 
time scales, decoherence is not a major constraint on predictability
for ``realistic'' coarse-grainings. Noise interferes with the 
predictability of the nonlocal coarse-grainings much more than
for the local ones. The equations of motion of the local
coarse-grainings require many fewer steps to evolve over a given time
to a given accuracy than do those for the nonlocal coarse-grainings.
In short, we find that the familiar local coarse-grainings are 
more predictable than the nonlocal ones.

\section{The Linear Chain}
\label{sec:II}
\subsection{ The Chain and Its Coarse-Grainings}
In this section we lay out the details of the model. We consider a linear
chain of ${\cal N}$ ``atoms'', each of mass $\mu$, separated at rest by equal
distances $\Delta x$, and with displacements
$x_i, i=0, \cdots, {\cal N}-1$. (See Figure 1.)
Each atom interacts linearly with its nearest
neighbors. The action describing this system is: 
\begin{equation}
S\left[x_j(\tau)\right] = \frac{1}{2}\, \mu\,
\sum^{{\cal N}-1}_{j=0}\ \int^{t_f}_0\, dt\, \left\{\left(\dot x_j(t)\right)^2 
-\omega^2
\left[x_{j+1}(t)-x_{j}(t)\right]^2\right\}
\label{twoone}
\end{equation}
where, using periodic boundary conditions, we take $x_{\cal N} = x_{0}$.

The linear harmonic chain is the simplest model of a solid and is treated
in many references.\footnote{For a classic reference see, \cite{Bri46}.}
The $x_j(t)$ may be conveniently analyzed in terms of spatial modes
\begin{equation}
x_j(t) = \sum^{{\cal N}/2}_{\ell=0}\ \left[a_\ell (t) f_\ell (j)
+ c.c.\right]
\label{twotwo}
\end{equation}
where
\begin{equation}
f_\ell(j) = {\cal N}^{-\frac{1}{2}}\, \exp (2\pi ij\ell/{\cal N})\ .
\label{twothree}
\end{equation}
These are the normal modes of the chain and the $a_\ell(t)$ oscillate with
frequencies
\begin{equation}
\omega_\ell = 2\omega\sin (\pi\ell/{\cal N})\ .
\label{twofour}
\end{equation}
The action expressed in terms of normal modes takes the simple form
\begin{equation}
S\left[a_\ell(\tau)\right] = \frac{1}{2}\, \mu\,
\sum^{{\cal N}/2}_{\ell=0}\ \int^{t_f}_0\, dt\, \left(|\dot
a_\ell(t)|^2 -\omega_\ell^2 |a_\ell(t)|^2\right) \ .
\label{twofive}
\end{equation} 

\begin{figure}[t]
\begin{center}
\epsfig{file=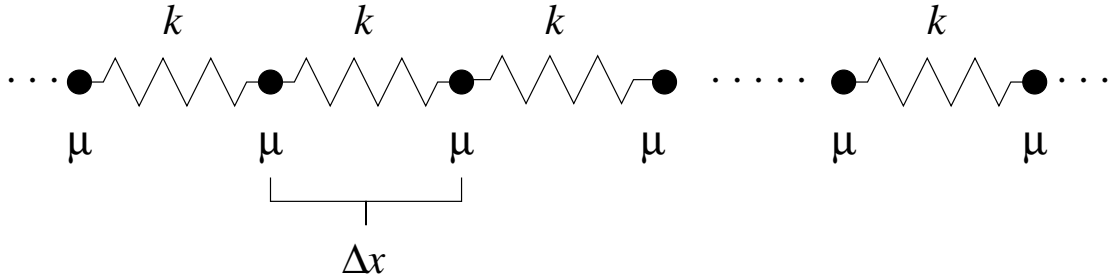}
\bigskip
\bigskip
\caption{The microscopic picture of the one-dimensional harmonic
chain.  ${\cal N}$ equal masses $\mu$ are spaced an equilibrium distance
$\Delta x$ apart.  The displacement of the $j$th mass is denoted
$x_j$. The masses have a linear restoring force between them with
a spring constant $k=\mu\omega^2$.  The chain is assumed to have periodic boundary conditions.}
\label{fig1}
\end{center}
\end{figure}

Both classically and quantum mechanically we will compare members of a
family of coarse-grainings defined in part by two integer parameters $N$ and $d$.
The ${\cal N}$ particles in the chain are divided into 
${\cal M}={\cal N}/N$ {\it
groups} of $N$ atoms each. Each group consists of $N/d$ {\it clumps} of
$d$ particles spaced by ${\cal M}d$. (See Figure 2.) 
When $d = N$ there is only one
clump. This coarse-graining is local---the atoms are as
closely spaced as possible in the
chain. When $d=1$ there is only one particle per clump. This coarse-graining
is nonlocal, as the atoms are as spread
out as possible in the chain. The variables
followed by the coarse-graining are the average displacements of the groups.
\begin{equation}
X^{(d)}_J (t) = \frac{1}{N}\ \sum^{N/d-1}_{k=0}\ \sum^{[d/2]}_{m = [-d/2]+1}\ 
\ x_{Jd + m + k{\cal M}d} (t)\ .
\label{twosix}
\end{equation}
The first sum is over the clumps in the group, and the second sum is
over the atoms in a clump. 
The range of the sum over $m$ depends on whether $d$ is even or odd. 
For odd $d$ it ranges from $-(d-1)/2$ to $(d-1)/2$. For even $d$
the range is $-d/2+1$ to $d/2$. The brackets $[x]$ in (\ref{twosix}) denote rounding down to
the next lower integer.  The range of $J$ is
$J=0, \cdots, {\cal M}-1$ making a total of ${\cal M}$ coarse-grained
coordinates.

We consider sets of histories constructed from exhaustive sets of
exclusive equal ranges $\Delta$ of $X_J^{(d)}$ separated 
in by time intervals $\Delta t$. There is therefore a family of
coarse-grainings each member of which is labeled by values of 
the four parameters $(N, d, \Delta, \Delta t)$. 
\begin{figure}[t]
\begin{center}
\epsfig{file=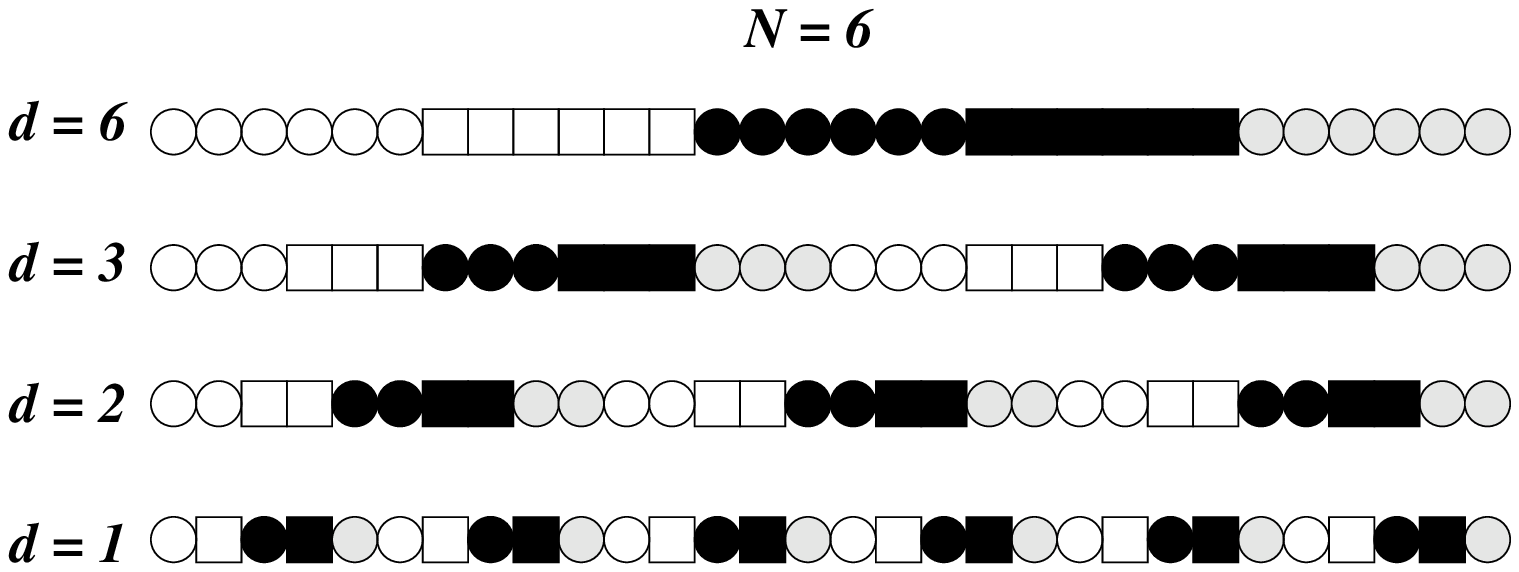}
\bigskip
\bigskip
\caption{The family of coarse-grainings under consideration.
The total number of masses ${\cal N}$ is divided into ${\cal M}$ groups
of $N$ masses each.  These groups are then further subdivided into
$N/d$ {\it clumps} of $d$ masses each. The clumps of a given group are
spaced so that clumps of all the other groups occur  before a clump is
repeated, that is, by a distance ${\cal M}d$. 
The coarse-grained variable $X_J$ is the average displacement
of the masses in the $J$th group.
In the figure, ${\cal N}=30$, ${\cal M}=5$, and $N=6$.
Masses in the same group are labeled with the same symbol, with a
different symbol (including the shading) for each group.  The arrangements are shown for
$d=1,2,3,6$, with $d=1$ being completely delocalized and $d=6$ being
completely localized.}
\end{center}
\label{fig2}
\end{figure}

It is convenient to expand the coarse-grained coordinates in terms of
spatial modes as well
\begin{equation}
X^{(d)}_J (t) = \sum^{{\cal M}/2}_{L=0}\ \left[A^{(d)}_L (t) F^{(d)}_L (J) +
c.c.\right]
\label{twoseven}
\end{equation}
where
\begin{equation}
F^{(d)}_L (J) = {\cal M}^{-1/2} \exp(2\pi iJL/{\cal M)}\ .
\label{twoeight}
\end{equation}
The parameter $L$ labels the coarse-grained mode $A^{(d)}_L$ just
as $\ell$ labels the fine-grained mode $a_\ell$.  The corresponding frequency
$\Omega^{(d)}_L$ depends on the choice of coarse-graining $d$.
The $A^{(d)}_L (t)$ will be a superposition of the normal modes of the
chain. Specifically,
\begin{equation}
A^{(d)}_L (t) = \sum^{d-1}_{k=0}\, c^{(d)}_{Lk}\, a_{m(k)N/d}(t) 
\label{twonine}
\end{equation}
where
\begin{equation}
m(k) = \cases{L+k{\cal M}/2, & $k$ even\ ,\cr
  -L+(k+1){\cal M}/2, & $k$ odd \ .  \cr}
\label{twoten}
\end{equation}
The coefficients $c_{Lk}$ are given by
\begin{equation}
c^{(d)}_{Lk} = \frac{1}{d\sqrt{N}}\ \ \frac{\sin\left[\pi m(k)/{\cal
M}\right]}{\sin\left[\pi m(k)/({\cal M}d)\right]}
\ e^{i m(k) \varphi(d)}
\label{twoeleven}
\end{equation}
with the phase $\varphi(d)$ vanishing if $d$ is odd and equal to
$-\pi/({\cal M}d)$ if $d$ is even.

In the delocalized limit, $d=1$, the $A^{(1)}_L$ are just single normal modes
\begin{equation}
A^{(1)}_L(t) \propto a_{LN}(t).
\label{twotwelve}
\end{equation}
By contrast, in the localized limit $d=N$, the $A^{(N)}_L(t)$ are
superpositions of all modes higher than $L$, {\it viz.}
\begin{equation}
A^{(N)}_L (t) = \sum^{N-1}_{k=0}\, c^{(d)}_{Lk}\, a_{m(k)}(t).
\label{twothirteen}
\end{equation}
These facts will be useful in what follows.

\subsection{Environments}

For several purposes in the subsequent analysis it will be convenient
to divide the configuration space of fine-grained  modes $\{a_\ell\}$
up into a subspace spanned by variables that are followed by the coarse-grainings introduced above,
and and a subspace spanned by variables that are ignored. The $\cal M$ variables $\{A^{(d)}_L\}$
may be said to define the configuration space of the ``system''. 
There is considerable latitude in the choice of variables
$\{q^{(d)}_a\}, a=1,2, \cdots, {\cal N} - {\cal M}$ that define
the ``environment''. The only requirement is that
the $\{A^{(d)}_L\}$ and the $\{q^{(d)}_a\}$ span the whole configuration
space of the fine-grained modes $\{a_\ell\}$. A convenient choice 
for our calculations will be to take the $q^{(d)}_a$ to be 
all but the lowest fine-grained modes that contribute to the $A^{(d)}_L$
through (\ref{twonine}). That is, we take the $q^{(d)}_a$ to be the 
real and imaginary parts of:
\begin{equation}
\label{twofourteen}
a_{m(k)N/d}, \, \, k=1,2, \cdots, d-1; \, \, L=0,
\cdots, {\cal M}/2 \ .
\end{equation}
Each label $a$ corresponds to a pair $(L,k)$; the exact form of the
correspondence will not be important for us. Thus defined, the 
environment variables are normal modes which classically obey: 
\begin{equation}
\label{twofifteen}
\ddot q_a^{(d)} + (\omega_a^{(d)})^2 q_a^{(d)} =0 \ . 
\end{equation}
The definition (\ref{twofourteen}) defines a different environment for each
$d$, and even for a given $d$ it is but one of many possible choices,
though the effect on the system must be independent of this choice.

Since only $d$ fine-grained modes contribute to each coarse-grained mode
$A^{(d)}_L$, this ``system plus environment''
only includes the full set of modes in
the localized case $d=N$.  For all other $d$, there will be some
fine-grained modes that contribute to none of the coarse-grained modes,
and which thus can be neglected.
When  $d=1$, 
each coarse-grained mode $A^{(1)}_L$ depends on only a single fine-grained
mode, and hence has no environment at all.

\subsection{ The Initial State}

As we discussed in the Introduction, the classical behavior of any 
quantum system is exhibited through the probabilities of {\it certain}
decoherent sets of alternative coarse-grained histories arising from 
{\it particular} initial conditions. In this paper we compare the 
predictability of the different members of the  family of
coarse-grainings introduced above for a single class of initial 
conditions. That is like the situation in cosmology, where the
initial condition is  given by fundamental theory and we search
among the possible sets of coarse-grained alternative histories
for sets that exhibit more predictability than other sets. 

In this  simple model of the harmonic chain there are no interactions
between different normal modes. This is an idealization;
realistically there will be interactions. These interactions mean
that the normal modes will tend to equilibrate. The highest frequency
modes can be expected to come into equilibrium first, the lowest 
frequency modes last. Thus, at an intermediate time in the 
approach to equilibrium, the chain can be described by a state
in which a range of low frequency normal modes are not yet in
equilibrium, but with higher frequency modes that are in equilibrium at
a temperature $T$.

Indeed, a garden variety string would be described in this way. 
In a plucked violin string, for instance, the wavelengths that
are reasonable fractions of the length of the string are not
in equilibrium, but very short wavelengths---short compared
to the transverse dimensions of the string, certainly---are
likely to be in equilibrium. 

Classically an initial condition is given by a distribution on
the phase-space of mode amplitudes $a_\ell$ and their conjugate
momenta $\pi_\ell$.  We denote these collectively by $\{z_\ell\} =
\{(a_\ell, \pi_\ell)\}, \ell = 1, 2, \cdots \cal N$.  We assume that the
significant expected values of $z_\ell$, denoted ${\bar z_\ell}$, 
occur only low frequencies, so that for $\ell$ greater than a cutoff  $ \ell_C$ we have
$z_\ell \approx 0$.  We assume thermal fluctuations around these expected
value. A distribution which gives this is:
\begin{equation}
\rho(\{z_\ell\})=\prod\limits_{\ell=0}^{{\cal N}-1}
(Z_\ell^{-1} e^{-H_\ell(z_\ell-{\bar z_\ell})/T}) \ .
\label{twosixteen}
\end{equation}
Here $H_\ell(z_\ell)$
is the Hamiltonian $(\pi_{\ell}^2/\mu +\mu\omega_{\ell}^2 a_\ell^2)/2$ for
each mode,  and $Z_\ell$ is a normalizing factor.

The corresponding assumption in quantum mechanics would be
a density matrix of the form:
\begin{equation}
\rho=\prod\limits_{\ell=0}^{{\cal N}-1}\,
\exp[i({\bar a_\ell}^*{\hat \pi}_\ell - {\bar \pi_\ell}^*{\hat a}_\ell
  + {\rm h.c.})/2]\,
  (Z_\ell^{-1} e^{-H_\ell/T})\,
  \exp[-i({\bar a_\ell}^*{\hat \pi}_\ell - {\bar \pi_\ell}^*{\hat a}_\ell
  + {\rm h.c.})/2]
\label{twoseventeen}
\end{equation}
where ${\hat a}_\ell$ and ${\hat\pi}_\ell$ are the quantum operators
corresponding to the classical variables $z_\ell = (a_\ell,\pi_\ell)$,
and the exponentials sandwiching the thermal state are phase-space
displacement operators.
These are the initial conditions we assume in our analysis.

A system is said to be in  thermal equilibrium when its state
is such that the probabilities of quasiclassical alternatives
are accurately reproduced by a thermal density matrix.
Typical examples are coarse-grainings by volume averages of densities
of conserved quantities such as energy, momentum, {\it etc.} However
the probabilities of an arbitrary set of alternatives will
not be generally be reproduced by the thermal density matrix. 
It is plausible that many (or perhaps most) initial states of
the linear chain relax after a suitable time to a state of
local thermal equilibrium for which the  
probabilities of the {\it local} coarse-grainings described above are 
reproduced by a density matrix of the form (\ref{twosixteen}) or
(\ref{twoseventeen}). However, we have assumed something much stronger
which is that the probabilities of the whole family of coarse grainings
are accurately reproduced by these density matrices. That is 
a much more restrictive condition on the initial condition.   

\section{Equations of Motion for Classical Coarse-Grainings}
\label{sec: III.}

In this section we consider the {\it classical} linear chain.
Each atom moves according to a classical equation of motion. However,
with a probabilistic initial condition, 
a set of coarse-grained variables (such as the $X^{(d)}_J$ discussed in the
previous section) need not obey {\it any} closed system of deterministic
equations. We analyze the constraints on the initial distribution
$\rho(z^0)$ necessary for the $\{X^{(d)}_J\}$ for each $d$ to obey and
closed set of deterministic equations and we exhibit the form these
equations take. The general kind of analysis we describe has been
considered by many authors, for example Zwanzig \cite{Zwa73} and Brun
\cite{Bru93}. 
What is new here is the application of these methods to comparing 
different coarse-grainings of
the linear chain.

\subsection{The Probability of Determinism}

A set of functionals of the phase-space paths $F_A[z(t)]$, $A=1,\cdots, P$
and a set of ranges $\{\Delta_\alpha\}$ in ${\bf R}^P$ define the most
general kind of classical coarse-graining. The probability $p_\alpha$ that
the functionals $F_A$ have values in the range $\Delta_\alpha$ is
\begin{equation}
p_\alpha = \int \delta z\, e_\alpha\left(F_A[z(t)]\right)\, \delta\,
[z(t) - z_t (z^0)]
\, \rho(z^0)
\label{threeone}
\end{equation}
Here, $e_\alpha$ is the characteristic function  (1 inside,
0 outside) for the range $\Delta_\alpha$  and $z_t(z^0)$ is the classical
evolution of the initial data $z^0$. 
The functional $\delta$-function
enforces that evolution, assigning zero probability to all paths which do not
conform to it.
The functional integral is 
over all phase-space paths $z(t)$ including an integral over the
initial conditions $z^0$. 

Utilizing this framework one can calculate the probability that the
coarse-grained variables such as the $\{X^{(d)}_J\}$ exhibit deterministic
correlations in time. The functionals $F_A$ for example might defined
different orbits in phase-space. Even more simply, one can evaluate the
probability that a set of deterministic equations of motion\footnote{We
follow the convention that $f(t, x(\tau)]$ means that $f$ is a function of
$t$ and a functional of $X(\tau)$, for example $\int^t_0 x(\tau) d\tau$.}
\begin{equation}
{\cal E}_J(t, X(\tau)]=0
\label{threetwo}
\end{equation}
holds for a set of variables $\{X^{(d)}_J\}$ 
(denoted without indices in (\ref{threetwo})) at time $t$, by calculating the probability that the
functionals ${\cal E}_J(t, X(\tau)]$ have the value zero. If that
probability is high then the coarse-grained variables obey the
deterministic set of equations (\ref{threetwo}).

\subsection{Equations of Motion}

We now consider equations of motion for the coarse-grained average
positions $X^{(d)}_J(t)$
introduced in Section II.  Equivalently, and more conveniently,
we can consider the equations of motion for the corresponding spatial modes $A^{(d)}_L(t)$ defined by
(\ref{twoseven}) and given terms of the fine-grained modes by
(\ref{twonine}). It is convenient to take the system-environment split
defined by (\ref{twofourteen}) in which case the ignored
coordinates are the $\{q^{(d)}_a\}$ of (\ref{twofifteen}).

The normal modes of the chain $a_\ell$ oscillate with the frequencies
$\omega_\ell$ given in eq (\ref{twofour}).  Evidently as a consequence of 
(\ref{twonine}):
\begin{equation}
\ddot A_L(t) = -\sum^{d-1}_{k=0} c_{Lk}\ \omega^2_{m(k)N/d}\ a_{m(k)N/d}(t)\ .
\label{threethree}
\end{equation}
We have suppressed the superscript $(d)$ in this equation, as we shall
do for clarity in the remainder of this section. 
The ignored coordinates were chosen to coincide with all but the
lowest frequency normal mode in (\ref{threethree}). Therefore that equation
may be rewritten as
\begin{equation}
\ddot A_L(t) + \Omega^2_L A_L(t) = f_L(t)\ ,
\label{threefour}
\end{equation}
where
\begin{equation}
\Omega_L = \omega_{LN/d}\ ,
\label{threefive}
\end{equation}
and
\begin{equation}
f_L(t) = \sum^{d-1}_{a=1} \left(\Omega^2_L - \omega^2_a\right)
c_{La}\ q_a(t)\ .
\label{threesix}
\end{equation}
Here, $\omega_a$, $c_{La}$, etc.~are the values of $\omega_\ell$, $c_{Lk}$
appropriate to the value $a$.

The simple equation of motion (\ref{twofifteen}) can be solved to 
express the $q_a(t)$ in terms of their initial values
\begin{equation}
q_a(t) = q_a(0)\, \cos\, (\omega_a t) + \frac{p_a(0)}{\omega_a}\, 
\sin\ (\omega_a t)\ .
\label{threeeight}
\end{equation}
The $q_a(t)$ are probabilistically distributed with a distribution that
depends on the distribution of the initial conditions through
(\ref{threeeight}). Let $F_L(t)$ be the expected value of $f_L(t)$,
\begin{equation}
F_L(t)= {\rm M}[f_L(t)]   
\label{threenine}
\end{equation}
where ${\rm M}[f_L]$ denotes the mean over the distribution of initial
conditions. (We use ${\rm M}[\cdot]$ for classical expected values and
$\langle\cdot\rangle$ for quantum mechanical ones.)
Then equation (\ref{threefour}) becomes
\begin{equation}
{\cal E}_L(t, A_L(\tau)] \equiv \ddot A_L(t) + \Omega^2_L A_L(t) - F_L(t) =
\Delta f_L(t) \ ,
\label{threeten}
\end{equation}
where
\begin{equation}
\Delta f_L(t) = f_L(t) - {\rm M}[f_L(t)]\ .
\label{threeeleven}
\end{equation}

We can think of ${\cal E}_L(t, A_L(\tau)]=0$ as a deterministic equation of
motion and $\Delta f_L$ as classical stochastic noise causing
deviations from that determinism. Since the expected value of ${\cal E}_L(t,
A_L(\tau)]$ is
zero,  the coarse-graining will approximately
obey the deterministic equation ${\cal
E}_L(t, A_L(\tau)]=0$ provided the noise is small. 
A simple measure of its size is
\begin{equation}
{\rm M}\left[\Delta f_L(t) \Delta f_L(t^\prime)\right]\ .
\label{threetwelve}
\end{equation}

The situation is even simpler if, in addition,  $F_L(t)$ is negligible, as it is
in many realistic situations. Then the chain obeys the equation
\begin{equation}
\ddot A_L(t) + \Omega^2_L A_L(t)= \Delta f_L(t) \approx 0
\label{threethirteen}
\end{equation}
with high probability---an equation whose form is independent of the
initial distribution.

For our chain model $F_L(t)$ does not vanish.  Because the
initial condition is diagonal in the fine-grained modes rather than the
$A_L$'s and $q_a$'s, there is a non-vanishing contribution to the
driving force $F_L$ proportional to $(A_L(0)-{\rm M}[A_L(0)])$.  However,
even in terms of $A_L$ and $q_a$, the initial condition is close to
diagonal for large $N$ and $\cal M$; thus, this term is small
compared both to the noise and the terms of the homogeneous equation of motion
(\ref{threethirteen}).  The time average of $F_L(t)$ also vanishes.
Hence, we can safely neglect it.  

Unlike effective classical equations for many systems, eq
(\ref{threethirteen}) exhibits neither nonlocality or dissipation. These
simplifying features can be traced to the conservation of the energies of
the individual normal modes. The environment---the $q_a$'s---are all
normal modes. Energy, therefore, cannot be exchanged between the $A_L$ and
the $q_a$ on average, that is between the system and its environment.
Defining the energy for a coarse-grained mode is somewhat ambiguous,
but simple definitions give an energy which fluctuates about a fixed
average.  This is special
to this linear system which has a high level of integrability.

The initial distribution $\rho(z^0_\ell)$ determines whether the
classical noise is small. The noise will be negligible when there is a
negligible probability for any significant initial excitation for the $q_a$.
Thus, for initial distributions that favor low frequency modes
below some cut-off value,  we expect low noise and deterministic behavior of
the $A_L(t)$. We shall return to this question in more detail in Section V.

\subsection{Deriving the equations of motion from the Lagrangian}

While the above derivation is perfectly correct, it is in some sense
a short-cut to a more standard procedure.  In this, we would substitute
the change of variables (\ref{twoseven}) and (\ref{twofourteen}) into the expression for the action
(\ref{twoone}), and derive the Euler-Lagrange equations for the new variables.
This is important in making contact later with the quantum case, since
that derivation proceeds from the classical Lagrangian.

The action (\ref{twoone}) may be expressed more compactly in matrix
notation writing $\vec a(t)$ for the vector with components $a_\ell (t)$.
Then
\begin{equation}
S\left[\vec a(\tau)\right] = \frac{1}{2}\ \int^{t_f}_0 dt\ \left[\left(\vec
a\, {\bbox{\mu}}\, \vec a\right) - \left(\vec a\, {\bbox\mu}^{\frac{1}{2}}
\, \bbox\omega^2
\, {\bbox\mu}^{\frac{1}{2}}\, \vec a\right)\right]
\label{threefourteen}
\end{equation}
We use a Hermitian inner product indicated only implicitly, so the
${\bf(}\vec
a\vec b{\bf)} = \sum_\ell a^*_\ell b_\ell$.
Sandwiched in inner products like those of (\ref{threefourteen}),
${\bbox \omega}^2$ and ${\bbox \mu}$ denote the diagonal matrices of normal mode
frequencies and masses (all equal) respectively.  For this case,
$\bbox\mu^{1/2}\bbox\omega^2\bbox\mu^{1/2}=\mu\bbox\omega^2$.

The coarse-grainings are by the variables $X^{(d)}_J$, or equivalently by
their modes $A^{(d)}_L$.  The coordinates $q^{(d)}_a$
are the fine-grained modes that are not followed
by the coarse-graining. Continuing to suppress superscript $(d)$, we write
\begin{equation}
a_\ell = \sum_L\, S_{\ell L} A_L + \sum_a\, T_{\ell a} q_a
\label{threefifteen}
\end{equation}
or in matrix notation
\begin{equation}
\vec a = {\bf S}\, \vec A + {\bf T}\, \vec q
\label{threesixteen}
\end{equation}
In general, there is considerable arbitrariness in the matrices ${\bf S}$ and ${\bf T}$ 
corresponding to how the variables $q_a$ are chosen. The only requirements are
that (\ref{threesixteen}) must reproduce the definition of the $A_L$'s
(\ref{twosix}) and (\ref{twoseven})  and the set of $(A_L, q_a)$ must span the space of the
$a_\ell$. The choice (\ref{twofourteen}) specifies a definite ${\bf T}$ and
${\bf S}$. 

Inserting the transformation (\ref{threesixteen}) into the action
(\ref{threefourteen}) one has
\begin{eqnarray}
S\left[\vec A(t), \vec q(t)\right] &=& \frac{1}{2}\ \int^{t_f}_0\, dt\,
\Biggl[\left(\dot{\vec A}\, {\bf M}_{SS}\, \dot{\vec A}\right)
+ \left(\dot{\vec q}\, {\bf M}_{TT}
\, \dot{\vec q}\right)\nonumber\\
&+&\left(\dot{\vec A}\, {\bf M}_{ST}\,\dot{\vec q}\right) + \left(\dot{\vec
q}\, {\bf M}_{TS}\, \dot{\vec A}\right) - \left(\vec A\, {\bf V}_{SS}\, \vec
A\right)\nonumber\\
&-&\left(\vec q\, {\bf V}_{TT}\, q\right) - \left(\vec A\, {\bf V}_{ST}\, \vec
q\right) - \left(\vec q\, {\bf V}_{TS}\, \vec A\right)\Biggr]
\label{threeseventeen}
\end{eqnarray}
where
\begin{mathletters}
\label{threeeighteen}
\begin{equation}
{\bf M}_{AB} = {\bf A}^{\dagger}\, {\bbox \mu}\, {\bf B}\ ,
\label{threeeighteena}
\end{equation}
and
\begin{equation}
{\bf V}_{AB} = {\bf A}^{\dagger} \bbox{\mu}^{\frac{1}{2}} \bf{\bbox\omega}^2
\bbox{\mu}^{\frac{1}{2}}{\bf B} \ . 
\label{threeeighteenb}
\end{equation}
\end{mathletters}

With the action in this form, it is a simple matter to derive the
equations of motion for ${\vec A}$ and ${\vec q}$:
\begin{mathletters}
\label{threenineteen}
\begin{eqnarray}
{\bf M}_{SS}{\ddot{\vec A}} + {\bf M}_{ST}{\ddot{\vec q}}
  &=& - {\bf V}_{SS}{\vec A} - {\bf V}_{ST}{\vec q} \ ,
\label{threenineteena} \\
{\bf M}_{TT}{\ddot{\vec q}} + {\bf M}_{TS}{\ddot{\vec A}}
  &=& - {\bf V}_{TT}{\vec q} - {\bf V}_{TS}{\vec A}\ .
\label{threenineteenb}
\end{eqnarray}
\end{mathletters}

Eq. (\ref{threenineteen}) looks complicated,
but in fact we can easily recover our earlier
result.  Because the $q$ variables are normal modes,
they obey the usual harmonic oscillator motion
(\ref{twofifteen}).
By plugging the second equation into the first, 
the equation becomes
\begin{equation}
({\bf M}_{SS} - {\bf M}_{ST}{\bf M}^{-1}_{TT}{\bf M}_{TS}){\ddot{\vec A}} +
({\bf V}_{SS} - {\bf M}_{ST}{\bf M}^{-1}_{TT}{\bf V}_{TS}){\vec A} =
- ({\bf V}_{ST} - {\bf M}_{ST}{\bf M}^{-1}_{TT}{\bf V}_{TT}){\vec q}.
\label{threetwenty}
\end{equation}
If we then plug in the simple harmonic solution (\ref{threeeight})
for the ignored modes ${\vec q}$ the equation reduces to the simple
(\ref{threethirteen}) above
multiplied by the diagonal matrix 
$({\bf M}_{SS} - {\bf M}_{ST}{\bf M}^{-1}_{TT}{\bf M}_{TS})$.
(See Appendix B for the proof of this.)

This is a slightly more complicated derivation of the equation of motion,
but it still relies on the shortcut of knowing the solution for the
motion of ${\vec q}$.  That will not be available in quantum
mechanics. Without making use of this, we could solve the
linear equation (\ref{threenineteenb}) above for ${\vec q}(t)$
and insert it into (\ref{threetwenty}).
The solution for ${\vec q}(t)$ is
\begin{eqnarray}
{\bf M}^{1/2}_{TT}{\vec q}(t) &=&
  \cos[{\bbox\Omega} t] {\bf M}^{1/2}_{TT}{\vec q}(0)
  + {\bbox\Omega}^{-1} \sin[{\bbox\Omega} t]
  {\bf M}^{1/2}_{TT}{\dot{\vec q}}(0) \nonumber\\
&& - {\bbox\Omega}^{-1} \int_0^t dt'\ \sin[{\bbox\Omega}(t-t')]
  {\bf M}^{-1/2}_{TT} ( {\bf M}_{TS}{\ddot{\vec A}}(t')
    + {\bf V}_{TS}{\vec A}(t')),
\label{threetwentyone}
\end{eqnarray}
where
\begin{equation}
{\bbox\Omega}^2 = {\bf M}^{-1/2}_{TT} {\bf V}_{TT} {\bf M}^{-1/2}_{TT},
\label{threetwentytwo}
\end{equation}
and it gives us the very complicated-looking equation of motion
\begin{eqnarray}
&& ({\bf M}_{SS} - {\bf M}_{ST}{\bf M}^{-1}_{TT}{\bf M}_{TS}){\ddot{\vec A}} +
({\bf V}_{SS} - {\bf M}_{ST}{\bf M}^{-1}_{TT}{\bf V}_{TS}){\vec A} \nonumber\\
&+& ({\bf V}_{ST} - {\bf M}_{ST}{\bf M}^{-1}_{TT}{\bf V}_{TT})
  {\bf M}^{-1/2}_{TT} {\bf \Omega}^{-1} \int_0^t dt'\ \sin[{\bf \Omega}(t-t')]
    {\bf M}^{-1/2}_{TT} ( {\bf M}_{TS}{\ddot{\vec A}}(t')
    + {\bf V}_{TS}{\vec A}(t')), \nonumber\\
&=& - ({\bf V}_{ST} - {\bf M}_{ST}{\bf M}^{-1}_{TT}{\bf V}_{TT})
   {\bf M}^{-1/2}_{TT} (\cos[{\bf \Omega} t] {\bf M}^{1/2}_{TT}{\vec q}(0)
  + {\bf \Omega}^{-1} \sin[{\bf \Omega} t] {\bf M}^{1/2}_{TT}{\dot{\vec q}}(0)).
\label{threetwentythree}
\end{eqnarray}

Two observations can be made.  First, the solution
(\ref{threetwentyone}) while still harmonic motion,
appears to be at quite different frequencies from the form
(\ref{threeeight}), and also includes a driving term absent
in that case.  Second, equation (\ref{threetwentythree})
appears very different from the rather simple harmonic oscillator
equation (\ref{threethirteen}).  These differences, however,
must be apparent rather than real; both (\ref{threeeight})
and (\ref{threetwentyone}) follow from the same Lagrangian, as do
(\ref{threethirteen}) and (\ref{threetwentythree}).
In fact, the more complicated form (\ref{threetwentythree})
of the equation of motion is related to the simple form
(\ref{threethirteen}) by an invertible transformation,
and thus has exactly the same solutions.  (See Appendix C.)

\subsection{Transforming the classical noise}

The fact that classical equations of motion can be represented in widely
different forms is nothing new, of course.  But the apparent complexity
masking the comparatively simple underlying dynamics
complicates the analysis of the quantum case.  For the classical
case, it is best to stick with the simple form (\ref{threethirteen}).

The two forms do differ in one important respect.
The retarded term in equation (\ref{threetwentythree}) is most readily
identified as part of the homogeneous equation of motion, while in
(\ref{threethirteen}) it is implicitly included
in the noise terms.  We shall see that this ambiguity in the definition
of the noise becomes important in comparing the classical and quantum
cases:  one definition is most natural in the classical case, while the
other is most natural in the quantum case (where ${\vec A}(t)$ need not
obey the classical equation of motion).

Let us restrict ourselves to a single coarse-grained mode $A_L^{(d)}(t)$
and its associated modes ${\vec q}(t)$.
If we define the two forms of the noise to be the right hand sides of
equations (\ref{threethirteen}) and (\ref{threetwentythree}), respectively,
they can be expressed
\begin{mathletters}
\label{threetwentyfour}
\begin{equation}
\Delta f_L(t) = {\vec c} \left(\Omega^2_L{\bf I} - {\bf\Omega}^2_Q\right)
  \left( \cos[{\bf\Omega}_Q t] \Delta{\vec q}(0) + {\bf\Omega}_Q^{-1}
  \sin[{\bf\Omega}_Q t] \Delta{\dot{\vec q}}(0) \right)\ ,
\label{threetwentyfoura}
\end{equation}
and
\begin{equation}
\Delta f_L'(t) = N
   {\vec c} \left(\Omega^2_L{\bf I} - {\bf\Omega}^2_Q \right)
   {\bf M}^{-1/2}_{TT} (\cos[{\bf\Omega} t] {\bf M}^{1/2}_{TT}\Delta{\vec q}(0)
  + {\bf\Omega}^{-1} \sin[{\bf\Omega} t]
  {\bf M}^{1/2}_{TT}\Delta{\dot{\vec q}}(0)),
\label{threetwentyfourb}
\end{equation}
\end{mathletters}
where $\vec c$ is a vector in the space of the $q$'s with elements
equal to the coefficients $c_b$, ${\bf\Omega}_Q$ is a diagonal matrix
on the space of the $q$'s with diagonal elements $\omega_b$, and
${\bf\Omega}$ is the effective
frequency matrix defined by (\ref{threetwentytwo}).
The structure of these two expressions is closely parallel, and in
Appendix C we show that we can switch from one form of the noise
to the other by means of an invertible linear transformation.
As far as determining the classical dynamics and predictability, they
are equivalent.

However, if we look at the absolute strength of the noise as a function
of $L$ or $d$, the form of the noise {\it can} make a difference.  We
assess this by looking at the correlation function (\ref{threetwelve})
for the two forms of the noise.

In the initial state (\ref{twosixteen}) we find the expectation values
\begin{eqnarray}
{\rm M}[\Delta q_b(0) \Delta q_{b'}^*(0)]
  &=& k_BT({\bf V}^{-1}_{TT})_{bb'},
  \nonumber\\
{\rm M}[\Delta {\dot q}_b(0) \Delta {\dot q}_{b'}^*(0)] &=&
  k_BT({\bf M}^{-1}_{TT})_{bb'}.
\label{threetwentyfive}
\end{eqnarray}
We can use these to calculate the correlation functions (\ref{threetwelve})
\begin{eqnarray}
{\rm M}\left[\Delta f_L(t) \Delta f_L(t^\prime)\right] &=& k_BT
  {\vec c} \left(\Omega^2_L{\bf I} - {\bf\Omega}^2_Q\right)
  \biggl(\cos[{\bf\Omega}_Q t] {\bf V}^{-1}_{TT}
  \cos[{\bf\Omega}_Q t'] \nonumber\\
&& +  {\bf\Omega}_Q^{-1} \sin[{\bf\Omega}_Q t]
  {\bf M}^{-1}_{TT} \sin[{\bf\Omega}_Q t']
  {\bf\Omega}_Q^{-1}  \biggr)
  \left(\Omega^2_L{\bf I} - {\bf\Omega}^2_Q\right) {\vec c}\ ,
\label{threetwentysix}
\end{eqnarray}
and
\begin{eqnarray}
{\rm M}\left[\Delta f'_L(t) \Delta f'_L(t^\prime)\right] &=& Nk_BT
  {\vec c} \left(\Omega^2_L{\bf I} - {\bf\Omega}^2_Q\right)
  {\bf M}^{-1/2}_{TT}
  \biggl(\cos[{\bf\Omega} t] {\bf\Omega}^{-2}
  \cos[{\bf\Omega} t'] \nonumber\\
&& +  {\bf\Omega}^{-1} \sin[{\bf\Omega} t] \sin[{\bf\Omega} t']
  {\bf\Omega}^{-1}  \biggr)
  {\bf M}^{-1/2}_{TT}
  \left(\Omega^2_L{\bf I} - {\bf\Omega}^2_Q\right) {\vec c}\ .
\label{threetwentyseven}
\end{eqnarray}

Using these expressions, we can estimate the mean strength of the noise
by taking the average over time,
\begin{equation}
S^2 = \lim_{t_f\rightarrow\infty} {1\over t_f} \int_0^{t_f}
  {\rm M}[(f_L(t) - F_L(t))^2] dt\ .
\end{equation}
In Fig.~3 we plot $S^2$ as a function of the coarse-graining $d$.  We
see that the noise strength falls off steeply as a function of $d$.
Thus, the noise becomes lower in the highly localized case, and the
motion of the localized coarse-graining is most predictable.
Later we shall see that the Lagrangian form of the noise is closely
related to the strength of decoherence in the quantum case,
and closely resembles this ``simple'' noise.

\begin{figure}[t]
\begin{center}
\input{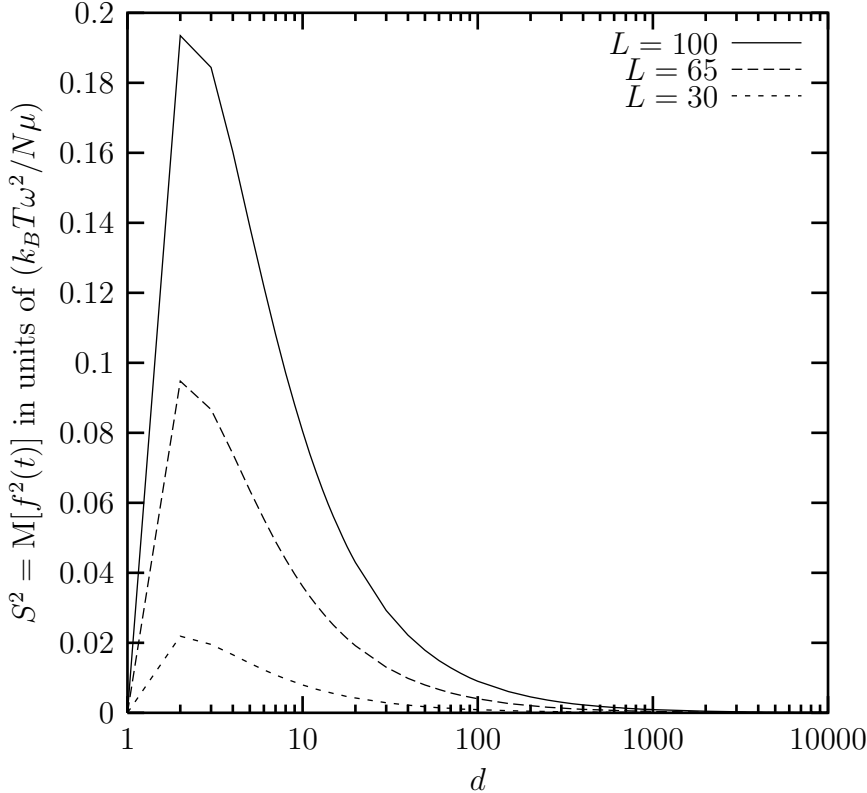}
\caption{The time-averaged mean force squared
of the noise, $S^2$, for the harmonic
chain as a function of the coarse-graining $d$.  This noise is chosen
for three typical coarse-grained modes $L=30,65,100$, with ${\cal M}=630$ and 
$N$ assumed to be very large; $N$ dependence is absorbed into the units,
$k_BT\omega^2/N\mu$.
Note that the noise vanishes for $d=1$, but otherwise assumes its highest
values at low $d$, dropping off like $1/d$ at high $d$, as is
predicted by the analytical result (\ref{fivesix}).}
\end{center}
\end{figure}

\section{The Wave Equation}
\label{sec: IV.}

The classical wave equation for the string does not follow directly from
the deterministic equations for the chain  (\ref{threethirteen}).  A restriction on the
initial distribution is required beyond that necessary for determinism.
This is the requirement that only very long wavelength modes of the chain
are excited. We assumed such a restriction on the initial distribution
in Section IIC, but in this section we will examine the requirements
on the short wavelength cutoff $\ell_C$. This  small gradient approximation is necessary for the validity
of the wave equation as for other familiar continuum equations such as the
Navier-Stokes equation.\footnote{See, e.g. \cite{LL59}, \cite{Zub74}.}

\subsection{The Small Gradient Approximation}

The expansion (\ref{twoseven}) of the coarse-grained coordinates $X^{(d)}_J(t)$ in terms of
coarse-grained modes $A^{(d)}_L(t)$ connects the equation of motion 
(\ref{threethirteen})
for the $A^{(d)}_L(t)$ to an equation of motion for the $X^{(d)}_J(t)$. The
character of these equations for the $X^{(d)}_J(t)$ is
determined by the dispersion relation for 
$\Omega^{(d)}_L$ which is [{\it cf.} (\ref{threefive}), (\ref{twofour})]:
\begin{equation}
\Omega^{(d)}_L = 2\omega\sin\left(\frac{\pi L}{{\cal M}d}\right)\ .
\label{fourone}
\end{equation}

Since (\ref{fourone}) is not linear in $L$, 
the equations for $X^{(d)}_J(t)$ following from (\ref{threethirteen}) will generally
relate $\ddot X^{(d)}_K(t)$ to all the other coordinates $X^{(d)}_J(t)$
of the chain.
The form of the equations for $X^{(d)}_J(t)$ simplifies if the initial
distribution is such that only modes with  
\begin{equation}
\ell \ll {\cal N}
\label{fourtwo}
\end{equation}
have any significant probabilities, that is $\ell_c \ll {\cal N}$. Then from (\ref{twofour}), $\omega_\ell
\approx 2\pi\omega\ell/{\cal N}$ and in particular
\begin{equation}
\Omega^{(d)}_L \approx \frac{2\pi\omega L}{{\cal M}d}\ . 
\label{fourthree}
\end{equation}
In this small gradient approximation, the equation of motion for the
$X^{(d)}_J(t)$ implied by (\ref{threethirteen})  and (\ref{twoseven}) is
\begin{equation}
\ddot X^{(d)}_J(t) = \frac{\omega^2}{d^2}\ \left[X^{(d)}_{J+1} (t) -
2X^{(d)}_J(t) + X^{(d)}_{J-1}(t)\right] \  .
\label{fourfour}
\end{equation}
Only nearest neighbor interactions are involved in the
small gradient approximations.

Usually a much stronger condition is meant by the small gradient
approximation,  namely that the only modes with significant probabilities
are those with
\begin{equation}
\ell \ll {\cal M} \ll {\cal N} \ ,
\label{fourfive}
\end{equation}
that is $\ell_C \ll {\cal M}$. 
For $d=N$ this condition  ensures that many groups will fit into
a wavelength so that the $X_J$ vary only slightly from one $J$ to the next.
We will see below that this is essential to deriving the continuum wave
equations.

Condition (\ref{fourfive}) can be very much stronger than (\ref{fourtwo}).
In a 10 cm length of string with typical interatomic spacings ${\cal N}
\sim 10^9$. Dividing the string into .1 mm lengths constituting the
groups in the $d=N$ case gives $N\sim 10^6$.

The condition (\ref{fourfive}) would imply that no values of $A^{(1)}_L(t)$
would be excited above $L=0$ since the $\ell$ values contributing to any
$d=1$ mode are all larger than $N$ [{\it cf.} (\ref{twonine})] except $\ell
= 0$. That makes the equation of motion (\ref{fourfour})
in the delocalized case not incorrect, but
rather trivial since both sides are negligible.

\subsection{The Continuum Approximation}

For the localized coarse-graining $d=N$, the difference equation
(\ref{fourfour}) is well approximated by the wave equation when $N$ is
large. The derivation is standard, but we briefly repeat its essential
features here.

Recall that we denoted the mass of an atom by ${\mu}$, and the spacing
between atoms in the unexcited chain by $\Delta x$. The mass density
$\sigma$ is therefore
\begin{equation}
\sigma = {\mu }/\Delta x \ . 
\label{foursix}
\end{equation}
Multiplying both sides of (\ref{fourfour}) by $\sigma$ it can be written
\begin{equation}
\sigma \ddot X^{(N)}_J (t) = {\mu}\omega^2\Delta x
\left[\frac{X^{(N)}_{J+1}(t) - 2X^{(N)}_J(t) + X^{(N)}_{J-1}(t)}{(N\Delta
x)^2}\right] \ .
\label{fourseven}
\end{equation}
The length $N\Delta x$ is the distance between the centers of the groups in
the unexcited string. Assuming the strong form of the slow approximation
gradient (\ref{fourfive}), we can approximate (\ref{fourseven}) by the
continuum equation
\begin{equation}
\sigma \frac{\partial^2X(x,t)}{\partial t^2} =
Y\frac{\partial^2X(x,t)}{\partial x^2}\ .
\label{foureight}
\end{equation}
where Young's modulus $Y$ is ${\mu}\omega^2\Delta x$. Eq (\ref{foureight})
is the wave equation for the propagation of compressional modes along the
string.

\section{Classical Predictability}
\label{sec:V}

The fine-grained variables of a classical system obey a closed system
of deterministic equations of motion. However, there is no guarantee 
that coarse-grained variables will. There may be no deterministic
equations at all, or the set of those that do hold may not close, 
in the sense that there may not be a complete set of equations to solve
for all the variables. 
In Section IIIB we demonstrated 
that the probabilities of the 
evolution of the coarse-grained modes $A^{(d)}_L(t)$ of the linear
harmonic chain are reproduced
by a classical equation of motion (\ref{threeten}) modified by noise:
\begin{equation}
\ddot A^{(d)}_L(t) - \Omega^2_L A^{(d)}_L(t) - F^{(d)}_L(t) =
\Delta f^{(d)}_L(t) \ 
\label{fiveone}
\end{equation}
for the family of coarse-grainings under consideration. 
The time evolution of the $A^{(d)}_L(t)$ will be classically predictable
by the left hand side of (\ref{fiveone}) if the noise term on the 
right hand side is negligible. In this section we analyze this
requirement for classical predictability as a function of $d$.

A simple estimate of the noise can be obtained by assuming a completely
thermal initial state characterized by a temperature $T$. 
That is, we assume (\ref{twosixteen}) with $\ell_C=0$. In this initial
state, ${\rm M}[f^{(d)}_L(t)] \approx 0$.
A measure of the magnitude of the noise
fluctuations is ${\rm M}[(f^{(d)}_L(t))^2]$. In a thermal state  
\begin{equation}
{\rm M}[a_{\ell}(t)a_{\ell'}(t)]=\delta_{\ell \ell'} k_BT/(2\mu\omega^2_\ell) \, .
\label{fivetwo}
\end{equation}
We therefore have from (\ref{threesix}):
\begin{equation}
{\rm M}[(f^{(d)}_L(t))^2]= \frac{k_BT}{\mu}\sum_{k=1}^{d-1} |c_{Lk}|^2 
\omega^2_{\ell(k)}\left(1-\frac{\omega^2_{\ell(0)}}{\omega^2_{\ell(k)}}
\right)^2 
  {\mu\over2}\left(\omega^2_{\ell(k)}{\bf V}_{TT}^{-1}
  + {\bf M}_{TT}^{-1}\right)_{kk}
\label{fivethree}
\end{equation}
where the final factor (arising from the initial condition) is very
close to $1$ for ${\cal M}\gg L$, that is,
$(\mu/2)(\omega^2_{\ell(k)}{\bf V}_{TT}^{-1}
+{\bf M}_{TT}^{-1})_{kk} = 1 + O(L^2/{\cal M}^2)$,
where $\ell(k) \equiv m(k)N/d$ is
\begin{mathletters}
\begin{eqnarray}
\ell(k)&=&(L+k{\cal M}/2)N/d \ , \quad {\rm for\ k\ even,} \label{fivefoura}\\
\ell(k)&=&(-L+(k+1){\cal M}/2)N/d \ , \quad {\rm for\ k\ odd} \label{fivefourb} \ . 
\end{eqnarray}
\end{mathletters}
Every term on the right hand side of (\ref{fivethree}) depends on $d$
although we have not indicated the dependence explicitly.

Figure 3 shows a plot of the expected value of the square of the 
fluctuations (\ref{fivethree}) as a function of $d$ for some 
representative values of $L$. The plot is for ${\cal M} = 630$;
the value of $N$ cancels out of the plotted function, but is assumed
to be very large in order to have sufficient factors $d$.  As is evident
from (\ref{fivethree}), the noise vanishes
for $d=1$. That is because, as we noted in (\ref{twotwelve}), $A^{(1)}_L$ is 
a single fine-grained mode  not coupled  to any other fine-grained
modes. This special situation results from our
idealized linear model of the chain. Even a tiny amount of nonlinearity
would couple this mode to others and produce noise.

Figure 3 shows that the noise generally increases with $L$ and 
decreases with $d$. Analytic estimates can be derived from 
(\ref{fivethree}) when $L\ll {\cal M}<N$. Assuming that the final factor
in (\ref{fivethree}) is well approximated by unity, we have for  $d=2$
\begin{equation}
{\rm M}[(f^{(d)}_L(t))^2]
  \approx \frac{k_BT\omega^2}{N\mu} (\pi L/{\cal M})^2 \ , 
\label{fivefive}
\end{equation}
while for large $d$ comparable to $N$
\begin{equation}
{\rm M}[(f^{(d)}_L(t))^2] \approx \frac{k_BT\omega^2}{N\mu}
  \frac{(\pi L/{\cal M})^2}{d} \ .
\label{fivesix}
\end{equation} 
The noise thus decreases inversely with $d$. For
the ``realistic'' coarse-graining discussed at the end of
Section IVA, where $N\sim 10^6$, the noise is $10^6$ times smaller
for $d=N$ than it is for $d=2$. That is a vast advantage in
predictability of the local coarse-grainings over the nonlocal
ones. The origin of this advantage can be traced to the approximate
conservation of the center of mass momentum of the local groups
of atoms.

The noise term on the right hand side of the equation of motion
(\ref{fiveone}) must be compared with the characteristic size of the
deterministic terms on the left hand side to get a true estimate
of the effect of noise on predictability. In (\ref{twosixteen}) we
assumed an initial state in which only fine-grained modes with
$\ell < \ell_C$ had significant excitations above thermal
noise. But, as we remarked earlier, the coarse-grained modes $A^{(d)}_L$
are superpositions of fine-grained modes with $\ell>LN/d$, as equation
(\ref{twonine}) shows. Thus, for $d<LN/\ell_C$, there will be no 
significant excitation of  $A^{(d)}_L$ above the level of thermal
noise. The subsequent dynamics is predictable only in a trivial 
sense. The string exhibits no motion except in response to thermal
fluctuations. For true classical predictability $d$ must be large, 
so the noise is low, but also so the deterministic terms in the 
equation of motion dominate the noise terms. 
We shall return to a more quantitative comparison of the noise and
dynamical force terms in the Conclusion. 

Even leaving aside the competition between noise and predictability,
the classical equations of motion for nonlocal coarse-grainings are 
distinguished from the local ones by their computational complexity.
Consider the classical equations (\ref{threethirteen}) for the
coarse-grained modes, or equivalently,  in the small gradient
approximation, eqs. (\ref{fourfour}) for the center of mass positions
of the groups. The characteristic dynamical time scales are of order
$d/ \omega$. The local coarse-grainings therefore vary the most slowly as would
be expected from their association with approximately conserved
quantities. Thus the computation of the evolution of the coarse-grained 
quantities $A^{(d)}_L$ or $X_J^{(d)}$ over a given time interval
to a given accuracy will take a factor of $N$ more time steps for
the nonlocal equations than the local ones. For the ``realistic'' case
when $N\sim 10^6$ that is a significant advantage in computational
complexity for the local coarse-grainings.

Thus, whether one considers the absolute value of the noise, the
relative size of the noise and deterministic forces, or the effort
needed to solve the classical equations of motion, the local
coarse-grainings are more predictable than the nonlocal ones in the
family we have considered. In the next section we turn to the same
questions in quantum theory.

\section{Decoherence}
\label{sec: VI}

\subsection{Quantum Mechanics of the Linear Chain}
In this section we turn to the classical behavior of the {\it quantum} linear
harmonic chain. A quantum system behaves classically when the probability
is high that coarse-grained histories exhibit the correlations implied by
classical equations of motion. We will be concerned  with the
histories of the linear harmonic chain coarse-grained by values of the
position averages $\{X^{(d)}_J(t)\}$ defined in Section II. More specifically we shall
consider, for each $d$, sets of histories defined by an exhaustive set of
exclusive regions  $\{{\bf\Delta}^k_{\alpha_k}\}$, $\alpha_k=1,2,3,\cdots$ of 
the $\cal M$-dimensional configuration space spanned by the 
$\{X^{(d)}_J(t)\}$ at a series of times $t_k,\ k=1,2,\cdots, n $ with 
$t_1<t_2\cdots<t_n$. We take the same set of regions for each value of
$d$ and usually take them to be ``cubes'' in $\cal M$ of equal sides
$\Delta$. 
An individual history is then labeled by the particular sequence of
intervals $(\alpha_1, \cdots, \alpha_k)\equiv \alpha$. We then compare the
probabilities that these histories are correlated by the classical
equations of motion discussed in Section III for different values of $d$.

Quantum interference between the 
individual members of a coarse grained set of histories must be negligible 
for probabilities to be consistently assigned to its individual members.
When this condition is satisfied the set of histories is
said to decohere. Decoherence is a prerequisite for probabilities.

The interference between histories $\alpha$ and $\alpha^\prime$ is measured
by the decoherence functional
\begin{equation}
D^{(d)}(\alpha^\prime,\alpha)
  = {\rm Tr}\left[P^d_{\alpha_n}(t_n) \cdots P^d_{\alpha_1}
(t_1)\, \rho\, P^d_{\alpha_1} (t_1) \cdots P^d_{\alpha_n} (t_n)\right]
\ .
\label{sixone}
\end{equation}
Here, $\{P^d_{\alpha_k}(t_k)\}$ are an exhaustive set of mutually
orthogonal Heisenberg picture projection operators projecting on the regions 
$\{{\bf\Delta}^k_{\alpha_k}\}$ of $\{X^{(d)}_J\}$ at time $t_k$. The operator $\rho$ is
the Heisenberg picture density matrix of the system. The set of histories
decoheres when the ``off-diagonal'' elements of $D^{(d)}$ are negligible,
\begin{equation}
D^{(d)}(\alpha^\prime, \alpha) \approx 0\ , \quad \alpha^\prime\not= \alpha\ .
\label{sixtwo}
\end{equation}

There is an equivalent path integral expression for the decoherence
functional of sets of histories coarse-grained by ranges of configuration
space such as those under discussion here. Suppose the coarse-graining is
entirely confined to times less than a final time $t_f$,
{\it i.e.},~$t_n<t_f$.  Then,
for the linear chain coarse-grained by ranges  of $X^{(d)}_J(t)$
\begin{eqnarray}
D^{(d)}(\alpha^\prime,\alpha)& = &\int_{\alpha^\prime} \delta a^\prime
\int_\alpha \delta a\, \delta \left(\vec a_f^\prime - \vec
a_f\right)\nonumber \\
&\times& \exp \left(i\left\{S \left[\vec a(t^\prime)\right] - S\left[\vec
a(t)\right]\right\}/\hbar\right)\, \rho \left(\vec a_0^\prime, \vec
a_0\right) \ .
\label{sixthree}
\end{eqnarray}
One integral in (\ref{sixthree}) is over
paths $\vec a(\tau)$ on the interval
$[0, t_f]$ which start at $\vec a_0$ at $t=0$ and
end at $\vec a_f$ at $t=t_f$ including integrations over those end points. 
The integral is only over paths which pass through
the regions  $\{{\bf\Delta}^k_{\alpha_k}\}$ in $\{X^{(d)}_J\}$ at the times $t_k$. The
constraint on the $X^{(d)}_J$ translates linearly into a constraint on the
$\vec a$ through (\ref{twotwo}).
An integral over $\vec a^\prime(\tau^\prime)$ is
similar except that it is constrained by the coarse-grainings of $\alpha^\prime$. 
The sum in (\ref{sixthree}) could have been expressed in terms of any 
configuration space variables. We have chosen the modes $a_\ell$ because
the action takes the simple form (\ref{twofive}).
Equally well, we could have used the coordinates of the individual atoms,
$x_i$, $i=1, \cdots, {\cal N}$. 

Path integrals of the form (\ref{sixthree}) have been extensively
studied for quadratic actions and thermal density matrices by many authors
\cite{FV63,CL83,GH93a,Bru93}. All integrals are Gaussian in this situation 
and can be evaluated explicitly. 
The simplest way to review this is to recall a
simple example.

\subsection{A Simple Example}

The sum-over-histories techniques used in this paper to calculate the
decoherence functional for sets of alternative coarse-grained histories of
the linear harmonic chain are generalizations of those
used for more straightforward coarse-grainings of
simpler linear systems \cite{GH93a,Bru93}.
In turn, these methods
extend those pioneered by Feynman and Vernon \cite{FV63}, and Caldeira and
Leggett \cite{CL83}.  While the results of this paper
are algebraically more complex, the basic
ideas are similar to those in simpler cases.
To emphasize this connection with previous work, and
to explain the ideas in an algebraically simple context, we review a
version of these simple models.

The model consists of the linear chain under discussion but with 
a particle of mass $M$ and position $X(t)$ coupled by a linear
interaction to atom 0. 
The total action is
\begin{equation}
S_{\rm tot} \left[X(\tau), \vec x(\tau)\right] = S\left[\vec x (\tau)\right]  
+ S_X [X(\tau)] + S_{\rm int} \left[x_0(\tau), X(\tau)\right]\ ,
\label{mone}
\end{equation}
where $S\left[\vec x (\tau)\right]$ is given by (\ref{twoone}), $S_X
[X(\tau)]$ is
\begin{equation}
S_X[X(\tau)] = \int^{t_f}_0 d\tau\ \frac{1}{2}\ M\dot X^2(\tau)
\label{mtwo}
\end{equation}
and the interaction term is 
\begin{equation}
S_{\rm int} \left[x_0(\tau) , X(\tau)\right]
 = - \kappa \int^{t_f}_0 d\tau\, X(\tau) x_0(\tau) \ . 
\label{mthree}
\end{equation}

We consider  coarse grainings  where only histories of $X$ are
followed. Thus the entire chain serves as the  
environment. 

The action $S[{\vec x}(\tau)]$ is
(\ref{twoone}), which is (\ref{twofive}) when expressed in terms of
normal modes. The interaction is 
\begin{equation}
S_{\rm int} \left[\vec a (\tau) , X(\tau\right] = - {\kappa} {\cal
N}^{-\frac{1}{2}} \sum_\ell \int^{t_f}_0 d\tau\, a_\ell (\tau) X(\tau) \
.
\label{mfour}
\end{equation}
The problem summarized by (\ref{mone}) can be mapped onto a problem
studied by many authors following Caldeira and Leggett \cite{CL83}. 
We will follow
the calculation of the decoherence functional described in \cite{GH93a} but
in a notation that is designed to stress the analogy with the subjects of
this paper. The translations are as follows, the first being the notation
of this paper, the second of \cite{GH93a}:
\begin{eqnarray}
a_\ell &\leftrightarrow& Q_A\ , \qquad \omega_\ell \leftrightarrow
\omega_A\ ,\nonumber\\
X(t) &\leftrightarrow& x(t)\ , \qquad \mu \leftrightarrow m\ ,\nonumber\\
{\kappa}{\cal N}^{-\frac{1}{2}}  
&\leftrightarrow& C_A\ , \qquad M \leftrightarrow M\ .
\label{mfive}
\end{eqnarray}

We assume an initial condition that is a product of a density matrix
$\tilde \rho(x^\prime_0\ ,\ x_0)$ for the particle and a thermal density
matrix $\rho_{T}(\vec a^\prime_0 , \vec a_0)$ at temperature $T$
for the environment. 

We consider a set of alternative coarse-grained histories defined by
exhaustive sets of ranges for $X$, 
$\left\{\Delta^1_\alpha\right\}\ , \
\left\{\Delta^2_\alpha\right\}\, \cdots \left\{\Delta^n_\alpha\right\}
\ , \quad \alpha=1, 2, \cdots$
at a series of times $t_1<\cdots<t_n$. The decoherence functional for this
coarse-grained set of histories is given generally by
\begin{eqnarray}
D(\alpha^\prime, \alpha) & = & \int_{\alpha^\prime} \delta X^\prime
\int_\alpha \delta X \int \delta a^\prime \int \delta a\, \delta
\left(X^\prime_f-X_f\right) \delta \left({\vec a'}_f - {\vec a}_f
\right)\nonumber\\
&\times& \exp \left(i\left\{S_{\rm tot}\left[X^\prime(\tau)\, ,
\, \vec a^\prime(\tau)\right] - S_{\rm tot} \left[X(\tau)\, ,
\, \vec a(\tau)\right]\right\}/\hbar\right)\nonumber\\
&\times& \tilde\rho \left(X^\prime_0\, ,\,  X_0\right)\ \rho_{T}
\left(\vec a^\prime_{ 0}\, ,\,  \vec a_{ 0}\right) \ . 
\label{msix}
\end{eqnarray}
The integral over the $X$'s is restricted to paths that traverse the
regions $\Delta^1_{\alpha_1},\cdots, \Delta^n_{\alpha_n}$ defining the
coarse-grained history $\alpha\equiv (\alpha_1, \cdots, \alpha_n)$ and
similarly for $X'$'s. The integrals over the $a^\prime_\ell$ and $a_\ell$
are unrestricted and,  given the assumed thermal initial condition, reduce
to Gaussian integrals which can be carried out explicitly. The result,
expressed in terms of the variables
\begin{mathletters}
\label{mseven}
\begin{eqnarray}
\bar X(t) &=& \frac{1}{2} \left[X^\prime(t) + X(t)\right] \ , \\
\label{mseven a}
\xi(t) &=& X^\prime(t) - X(t) \ , 
\label{mseven b}
\end{eqnarray}
\end{mathletters}
is
\begin{equation}
D\left(\alpha^\prime, \alpha\right) = \int_{\alpha^\prime} \delta x^\prime
\int_\alpha \delta x\ e^{i{\cal A}\left[\bar X(\tau),\ \xi(\tau)\right]} 
\tilde\rho
\left(\bar X_0 + \frac{\xi_0}{2}\ ,\ \bar X_0 - \frac{\xi_0}{2}\right)
\label{meight}
\end{equation}
where
\begin{eqnarray}
{\cal A}\left[\bar X(\tau)\, ,\, \xi(\tau)\right] &=& -\xi_0 M\left(d\bar
X/dt\right)|_{t=0} + \int^{t_f}_0 dt\, \xi(t) e(t\, ,\, \bar X(t)]\nonumber\\
&+& \frac{i}{4}\ \int^{t_f}_0 dt \int^{t_f}_0 dt^\prime\,  \xi(t^\prime) k_I
(t^\prime-t)\, \xi (t)\ .
\label{mnine}
\end{eqnarray}
The ingredients in eq.~(\ref{mnine}) are the equation of motion
\begin{equation}
e(t, X(\tau)] = - M{d^2 \bar X \over dt^2}(t) + \int^t_0 dt^\prime k_R
(t-t^\prime) \bar X(t^\prime) \ , 
\label{mten}
\end{equation}
together with the kernels
\begin{equation}
k_R(t) = - \kappa^2 \left(\mu{\cal N}\right)^{-1} \sum_\ell
\omega^{-1}_\ell \sin(\omega_\ell t)\ ,
\label{meleven}
\end{equation}
and
\begin{equation}
k_I(t) = \kappa^2 (\mu{\cal N})^{-1} \sum_\ell 
\omega^{-1}_\ell \coth 
\left(\frac{\hbar\omega_\ell}{2kT}\right) \cos \left(\omega_\ell t\right)
\ .
\label{mtwelve}
\end{equation}
The imaginary term in (\ref{mnine}) favors $\xi(t)=0$, that is $X^\prime
(t) = X(t)$. If it is large, the off-diagonal elements of $D(\alpha^\prime, \alpha)$
will be negligible and decoherence achieved. The characteristic time scale
$t_{\rm decoh}$ over which  enough
imaginary exponent is built up in between to make off-diagonal elements of
$D$ negligible is
\begin{equation}
t_{\rm decoh} \sim \left(k_I\Delta^2\right)^{-\frac{1}{2}}\ ,
\label{mthirteen}
\end{equation}
where $k_I$ and $\Delta$ denote characteristic sizes
of the kernel (\ref{mtwelve}) and the intervals $\Delta^i_{\alpha_i}$.

If the set of histories decoheres, then the restrictions of the coarse
graining on the integrals over $\xi(t)$ in (\ref{meight}) defining the
diagonal elements of $D(\alpha^\prime, \alpha)$ can be ignored and the
resulting Gaussian integral over $\xi(t)$ carried out. The result is the
following expression for the probabilities of the coarse-grained history
$\alpha$:
\begin{eqnarray}
p(\alpha) &=& \int_\alpha \delta \bar X \left[{\rm det}
\ k_I/4\pi\right]^{-\frac{1}{2}}\nonumber\\
&\times \exp&\left\{-\frac{1}{\hbar} \int^{t_f}_0 dt^\prime \int^{t_f}_0 dt\, 
 e (t^\prime,\bar X(\tau)]
k^{\rm inv}_I (t^\prime-t)\, e (t, \bar X(\tau)]\right\}
w(X_0, P_0) \ . 
\label{mfourteen}
\end{eqnarray}
Here, $k^{\rm inv}_I(t,t^\prime)$ is the inverse of $k_I(t, t^\prime)$ and
$w(X_0, P_0)$ is the Wigner distribution of the density matrix $\tilde
\rho$.

Eq.~(\ref{mfourteen}) shows that probabilities peak on histories obeying
the deterministic equation of motion $e(t, X(\tau)]=0$, but with a width in
this distribution related to $k_I(t^\prime-t)$. The same probabilities
follow from a Langevin equation
\begin{equation}
e(t, X(\tau)] + \ell(t)=0 \ , 
\label{mfifteen}
\end{equation}
with a stochastic noise force distributed with a correlation function
\begin{equation}
{\rm M}\left[\ell(t^\prime) \ell(t)\right] = \frac{1}{2} \hbar k_I(t^\prime-t)
\label{msixteen}
\end{equation}
Thus, decoherence and noise are connected. The stronger the coupling
between system and environment the more rapidly interference between
histories is dissipated, but also more noise which disturbs   
the deterministic dynamics and reduces predictability.

The deterministic equation $e(t,X(\tau)]=0$ is exactly
the same as the classical equation of motion which would be derived from
the Lagrangian for this system, in the limit of zero noise.  This
result is generally true for systems with quadratic Lagrangians, and
may hold approximately for systems with greater nonlinearities.
More details than we have given here are presented in \cite{GH93a}. We will
follow the basic ideas of this model in our treatment of the family of
coarse grainings of the linear chain.

\subsection{Decoherence Functional for the Linear Chain}

We now apply essentially the same procedure to the linear chain,
analyzing decoherence and deriving equations of motion for the
coarse-grained position averages $X^{(d)}_J$. 

Using the initial condition
(\ref{twoseventeen}) the integrals
over $q_a$ can be carried out in (\ref{sixthree}) yielding an expression for the
decoherence functional of the coarse-grained modes, $D^{(d)}[\vec A'(\tau),
\vec A(\tau)]$ analogous to (\ref{msix}). The magnitude of $D^{(d)}$ is 
\begin{eqnarray}
\left|D^{(d)}\left[\vec A'(\tau), \vec A(\tau)\right]\right| &=& \exp
  \Bigg[-\frac{1}{2}\ \int^{t_f}_0 dt^\prime \int^{t_f}_0 dt
  \left(\vec \xi (t^\prime) {\bf K}^{(d)}_I (t^\prime, t)\, \vec
  \xi(t)\right)\Bigg]\, \nonumber\\
&& \times \rho \left(\vec{\cal A}_0 + 
  \frac{\vec\xi_0}{2}, \vec{\cal A}_0
  - \frac{\vec\xi_0}{2}\right)
\label{sixtwentyone}
\end{eqnarray}
where
\begin{mathletters}
\label{sixtwentytwo}
\begin{eqnarray}
\vec\xi(t) &=& \vec A^\prime(t) - \vec A(t) \ ,
\label{sixtwentytwoa}\\
{\cal A}(t) &=& \frac{1}{2}\, \left[\vec A^\prime(t) + \vec A(t)\right]
\ .
\label{sixtwentytwob}
\end{eqnarray}
\end{mathletters}
The all-important kernel ${\bf K}^{(d)}_I (t^\prime, t)$ is given by
\begin{eqnarray}
{\bf K}^{(d)}_I(t',t) &=& {N^2 \mu^2 k_BT\over4\hbar^2}
  {\vec c} \left(\Omega^2_L{\bf I} - {\bbox\Omega}^2_Q\right)
  {\bf M}^{-1/2}_{TT} {\bbox\Omega}^{-1}
  \biggl(\cos[{\bbox\Omega} t] \cos[{\bbox\Omega} t'] \nonumber\\
&& + \sin[{\bbox\Omega} t] \sin[{\bbox\Omega} t']
  \biggr) {\bbox\Omega}^{-1} {\bf M}^{-1/2}_{TT}
  \left(\Omega^2_L{\bf I} - {\bbox\Omega}^2_Q\right) {\vec c}\ .
\label{sixtwentythree}
\end{eqnarray}
which is proportional to the correlation function (\ref{threetwentyseven})
of the classical noise derived from the Lagrangian.
Thus the ``off-diagonal'' elements of the decoherence functional 
decay exponentially with
increasing difference between $\vec A^\prime(\tau)$ and $\vec A(\tau)$.
Decoherence of coarse-grainings with suitably large regions
${\bf\Delta}^k_{\alpha_k}$ at suitably spaced intervals of time is thus
achieved. We shall return to a detailed discussion of decoherence times as
a function of $d$ below, but first we consider the equations of motion.

The probability of a given coarse-grained history $\alpha$ is given by
\begin{equation}
p(\alpha)= \int_\alpha\, \delta\vec A\,
D\left[\vec A(\tau), \vec A(\tau)\right]
\label{sixtwentyfour}
\end{equation}
which in analogy to (\ref{mfourteen}) can be expressed as
\begin{eqnarray}
p(\alpha) &=& \int_\alpha\, \delta\vec A\ \left[{\rm det}\,
\left({\bf K}^{(d)}_I\right)/4\pi\right]^{-\frac{1}{2}}\nonumber\\
&\times& \exp \Bigl\{-\frac{1}{\hbar}\ \int^{t_f}_0\, dt^\prime\, \int^{t_f}_0\,
dt\, \Bigl(\vec{\cal E}'(t^\prime, \vec A(\tau)] 
{\bf K}_I^{(d)\, {\rm inv}} (t^\prime, t) {\vec{\cal E}'}(t, \vec
A(\tau)]\Bigr)\Bigr\}\ .
\label{sixtwentyfive}
\end{eqnarray}
$\vec{\cal E}'(t, \vec A(\tau)]$ is a linear functional of $\vec A(\tau)$
analogous to (\ref{mten}),
whose exact form we return to below; ${\bf K}^{(d)\, {\rm
inv}} (t^\prime, t)$ is the inverse of ${\bf K}^{(d)}(t^\prime,t)$ defined in
(\ref{sixtwentythree}).

Eq (\ref{sixtwentyfive}) shows that the probability of histories is sharply
peaked about those obeying  the equations of motion
\begin{equation}
\vec{\cal E}'(t, \vec A(\tau)] = 0 \ , 
\label{sixtwentysix}
\end{equation}
but with Gaussian noise causing deviations from this predictability related to
the size of the kernel ${\bf K}_I^{(d)\,{\rm inv}} (t^\prime, t)$.
The form of the
equations of motion (\ref{sixtwentysix}) could be worked out by following
the procedure used in the simple example above. However, it is more direct
to note that the result of that calculation is exactly the equation of
motion that would be obtained by a completely classical analysis of the
coarse-grained dynamics, just as in (\ref{mfourteen}) for the simple model.
{\it The equations of motion} $\vec{\cal E}'(t, X(\tau)]=0$ {\it are
therefore equivalent to the classical equations} derived in Section III and
IV. This is an important simplification because a straightforward
extension of the above analysis would lead to nonlocal terms such as occur
in (\ref{mten}), as we saw in section IIIC.
In the end, these all cancel to give the simple equations of
motion exhibited in Section III.   This is checked explicitly in Appendix C.


The size of the kernel ${\bf K}^{(d)}_I(t, t^\prime)$ in (\ref{sixtwentyone}) 
controls the efficacy of decoherence---the larger the kernel the
shorter the decoherence time scale. 
 We study it as a function of $d$ holding all other
parameters fixed, including the temperature of bath $T$.
The kernel is necessarily positive \cite{Bru93}. A
simple measure of its size is the time-averaged trace
\begin{equation}
{\cal K}_I (d) = \lim_{t_f\rightarrow\infty} {1\over t_f}
  \int^{t_f}_0\, dt\ {\rm Tr}\left({\bf K}^{(d)} (t, t)\right)\ .
\label{sixtwentyseven}
\end{equation}

Figure 4 shows ${\cal K}_I$ plotted as a function of $d$.
Since the form of the kernel is closely related to the classical noise
correlation function ${\rm M}[\Delta f'(t)\Delta f'(t')]$ from
equation (\ref{threetwentyseven}),
it is not surprising that the shape of the result is virtually
identical to Figure 3.
While Figure 3 was computed from the form (\ref{threetwentysix})
for the noise, which is different, the difference amounts to a constant
factor of $N^2$ and a $d$-dependent factor very close to $1$.
{F}rom the two graphs,
where the $N$ dependence has been absorbed into the choice of units,
it is evident that the dependence on $d$ of the two
forms of noise is almost exactly the same.

\begin{figure}[htbp]
\begin{center}
\input{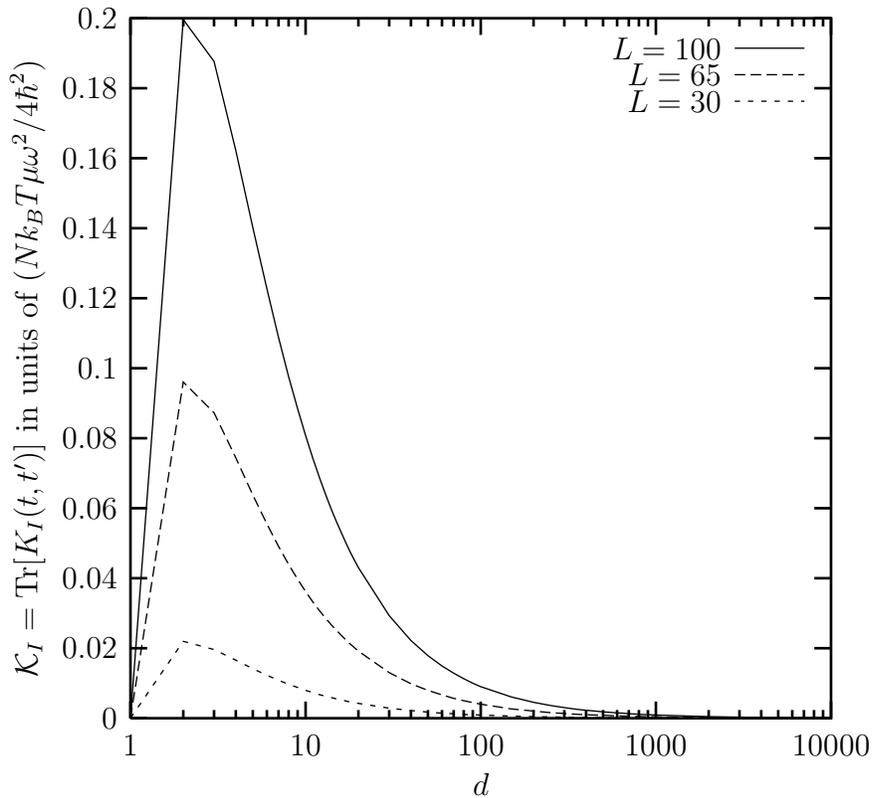}
\caption{The time-averaged trace ${\cal K}_I(d)$ of the decoherence
kernel ${\bf K}^{(d)}_I(t,t)$ in units of $Nk_BT\mu\omega^2/4\hbar^2$,
vs. the coarse-graining $d$,
for ${\cal M}=630$ and three $L$ values $L=30,65,100$.
The form of these curves is
virtually identical to the classical noise correlation function,
plotted in Figure 3.  Note that an overall factor of $N^2$ has been
divided out to make comparison to Figure 3 more exact. }
\end{center}
\end{figure}
Since the kernel  ${\bf  K}_I(d)$ is a factor of $N^2$
larger than the classical noise correlation function,
increasing the level of coarse-graining
$N$ exponentially improves decoherence, but actually reduces
the classical noise [{\it cf.} (\ref{fivesix})].  Because of this, for
realistically large values of $N$, the decoherence rate is rapid
compared to dynamical time scales
even in the completely localized case, where the
absolute strength of the noise is weakest.
One can understand this as increasing the absolute level of decoherence,
but simultaneously increasing the inertia of the coarse-grained variables
to resist the increased noise, since
the inertia of a group just goes like its total mass $N\mu$.
We discuss this trade-off more thoroughly below.

\section{Conclusions---Classicality}

For a system with many degrees of freedom, useful dynamical
predictions  concern regularities emerging from
coarse-grained descriptions.  To be sure, at a fine-grained level 
the system will display the
regularities arising from its fundamental equations of motion if
it is classical, or from the Schr\"odinger equation if 
quantum mechanical. However, these regularities are usually
impossible to extract or apply if the number of degrees of freedom
is very large. The useful predictions arise from much smaller
numbers of coarse-grained variables correlated in time according
to phenomenological equations of motion.

There are arbitrarily many sets of alternative
coarse-grained histories that can be assigned probabilities
on the basis of a closed system's initial condition and fundamental
dynamics. Which of these will exhibit useful regularities in time
governed by phenomenological equations of motion? How much, 
and what, coarse-graining is needed to obtain useful predictability?
This, very roughly, is the problem of characterizing classicality
that we mentioned in the Introduction. In this conclusion we
describe how our results for the linear harmonic chain bear
on this question.

Which coarse grained descriptions are predictable is an important problem
even in classical physics. However, it is especially important in
quantum mechanics where probabilities can be assigned only
to {\it decoherent} sets of histories, and two such sets
are generally mutually incompatible.
Further, in a loose sense, the number of sets of coarse-grained
histories is much larger in quantum mechanics than it is 
in classical physics. 
It is therefore important to explain {\it within} quantum
mechanics (and a theory of the cosmological initial condition) why
we find it useful to employ only a narrow class of the possible
coarse-grainings
by which the universe could in principle be described.  
Many see this as the central problem in understanding quantum
mechanics.

Characterizing classicality involves comparing the utility
for prediction of various sets of coarse-grained alternative
histories. In this paper we have considered only a family
of coarse-grainings of the simplest linear system exhibiting
a continuum description of the kind usually found in classical
physics---the linear harmonic chain.

Four parameters characterize the family of coarse-grainings we have considered.
The atoms of the chain are divided into groups of $N$, consisting
of equally spaced clumps of $d$ neighboring atoms. (See Figure 2.) The center
of mass coordinates of each group are coarse-grained by exhaustive
sets of equal ranges $\Delta$ spaced by equal time intervals $\Delta t$.
The four parameters are therefore $N, d, \Delta$ and $\Delta t$.

The probabilities of the individual members of a decoherent set
of histories follow from the initial state, so that any 
comparison of the predictability of different coarse-grainings
will depend crucially on the nature of this initial condition. We assumed
an initial condition in which the short wavelength modes of the
string were in thermal equilibrium at a temperature $T$, while
the long wavelength modes were excited well above the level
of thermal fluctuations. This is an initial state of local
but not global thermal equilibrium.

Both classically and quantum mechanically the probabilities of these
decoherent sets of alternative coarse-grained histories can be characterized as
arising from equations of motion augmented by noise. The dynamical 
time scale of the equations of motion (\ref{threethirteen}) is given 
roughly by
\begin{equation}
t_{dyn} \sim {d\over\omega}\left({{\cal M}\over L}\right)  \ . 
\label{sevenone}
\end{equation}
The local coarse-grainings $d \sim N$ have the longest dynamical time
scales because they  correspond to the approximately conserved center
of mass momenta of the groups. 

The utility of these equations of motion for
prediction depends on the size of the deviations from the regularity in 
time  they describe that is caused by the noise. The noise forces were
estimated roughly in eq. (\ref{fivesix}). (The force is $N\mu$ times
the force per unit mass $\Delta f^{(d)}_L$). 
For large $N$, $d$ comparable
to $N$, and $L$ small compared to ${\cal M}$:
\begin{equation}
F_{noise} \sim (k_BT \omega^2 \mu)^{1/2} (N/d)^{1/2}(L/{\cal M})\ .
\label{seventwo}
\end{equation}

With these basic estimates in hand, we can compare the different
members of  this family of coarse-grainings with respect to three
properties bearing on classicality:  the rate of decoherence,
the deviations from 
predictability caused by the noise, and the computational complexity
required to use the equations of motion to make predictions. 
The results are as follows:

$\bullet$ {\it Decoherence:}  Decoherence and noise are connected. 
The kernel which governs the size of the imaginary part of the
influence phase and effects decoherence is the correlation function
of the noise force [{\it cf.} (\ref{mtwelve}) and (\ref{msixteen})].  
{F}rom (\ref{sixtwentyone}) we can obtain the following 
rough estimate for the decoherence time $t_{decoh}$ which 
$\Delta t$ must exceed if the coarse-grained set of 
decoherent histories is to decohere:
\begin{equation}
t_{decoh} \sim \hbar/(F_{noise}\Delta) \ .
\label{seventhree}
\end{equation}
This time increases with the locality of the coarse-graining $d$
because $F_{noise}$ decreases with $d$, as (\ref{seventwo}) shows.  Since
the more local coarse-grainings are dominated by low frequency modes,
the dynamical timescale also increases with $d$, so that the ratio
of the two times favors the more local coarse-grainings;
combining (\ref{sevenone}) and (\ref{seventwo}) we have
\begin{equation}
\frac{t_{decoh}}{t_{dyn}} \sim (\frac{\lambda_{DB}}{\Delta})
(Nd)^{-1/2}\ ,
\label{sevenfour}
\end{equation}
where $\lambda_{DB}$ is the thermal De Broglie wavelength introduced
by Zurek \cite{Zur84}
\begin{equation}
\lambda_{DB} = \hbar/(k_BT \mu)^{1/2}  \ .
\label{sevenfive}
\end{equation}
The decoherence time scale must be less than the dynamical time scale
to use the equations of motion at all. This can be achieved by taking
$N$ or $\Delta$ or both sufficiently large, and for both local and
non-local coarse-grainings is not a very constraining 
condition given ``realistic'' parameter values,
as was stressed by Zurek \cite{Zur84} in simpler cases.
For $d \sim 1$, $N \sim 10^6, \mu \sim 10 AMU$, and $T \sim 300^\circ$,
\begin{equation}
t_{decoh}/t_{dyn} \sim (10^{-13} {\rm cm} /\Delta)  \ .
\label{sevensix}
\end{equation}
For such coarse-grainings,  decoherence is not a major restriction on
predictability.

$\bullet$ {\it Noise:} As (\ref{seventwo}) shows, the noise
force decreases as $d$ increases for fixed $N$, that is, it decreases as the
coarse-graining becomes more local.  However, it is
not the absolute scale of the noise that is important for
predictability, but rather its size relative to the dynamical force terms
$F_{dyn}$ that occur in the equation of motion. If $\cal L$ is the
characteristic size of the excitations of the chain that occur in the coarse
graining, then roughly 
\begin{equation}
F_{dyn} \sim  N \mu {\cal L}/t_{dyn}^2 \ .
\label{sevenseven}
\end{equation}
The size of $\cal L$ is determined by the initial condition and varies
with both $N$ and $d$. In (\ref{twosixteen}) and (\ref{twoseventeen})
we assumed an initial condition in which the fine-grained modes 
above a mode number $\ell_C$ were thermally excited, while modes
below $\ell_C$ were much more highly excited. 
The connection between fine- and coarse-grained modes given by
(\ref{twoseven}) and (\ref{twonine}) shows that the fine-grained
modes below $\ell_C$ contribute to coarse-grained modes only
when $d>N(L/\ell_C)$. Thus, if $d$ lies much below $N$,  we have
\begin{equation}
\frac{F_{noise}}{F_{dyn}} \sim 1,  \quad  d<<N \ ,
\label{seveneight}
\end{equation}
and the regularities of the equation of motion will be swamped by
the noise. 

By contrast, when $d \sim N$
\begin{equation}
\frac{F_{noise}}{F_{dyn}} \sim \frac{{\cal L}_T}{\cal L}\sqrt{N},
  \quad d \sim N \, 
\label{sevennine}
\end{equation}
where ${\cal L}_T$ is the characteristic scale of thermal excitations
of the mode:
\begin{equation}
{\cal L}_T \sim (k_BT/N\mu \Omega_L^2)^{1/2}  \ .
\label{seventen}
\end{equation}
If the size of excitations of long wavelength modes is much greater 
than that of thermal fluctuations, the effect of the noise on the equation
of motion will be negligible. Thus the local coarse-grainings exhibit
more regularity in time if the initial condition has this property.
For the realistic string described above with $N \sim 10^6$, typical thermal
excitations of the $L=10$ mode would be ${\cal L}_T \sim 10^-7$ m.

$\bullet$ {\it Computational Complexity:} Even when coarse-grainings
decohere so that probabilities can be assigned to histories, even when
noise is negligible so that histories exhibit the regularities
in time summarized by classical equations of motion, coarse-grainings 
can be distinguished by the effort needed to calculate these
regularities. The number of operations $N_S$ necessary to evolve the
equations of motion over a time interval $\cal T$ is roughly proportional
to
\begin{equation}
N_S \propto ({\cal M}/t_{dyn}){\cal T} \sim (L\omega{\cal T}/d) \ .
\label{seveneleven}
\end{equation}
Thus prediction becomes easier as both $N$ and $d$ increase. For 
``realistic'' coarse-grainings where $N \sim 10^6$ there  can be 
a significant difference between the local coarse-grainings with
$d \sim 10^6$ and the nonlocal ones with $d \sim$ few.

In summary, {\it decoherence, resistance
to noise and computational simplicity all favor local coarse-grainings
over nonlocal ones}---all these comparisons being contingent
on large $N$ and an initial condition in which long wavelength modes are more
excited than short wavelength ones. 

These quantitative results for the linear harmonic chain support the 
heuristic arguments for
the predictability of more general kinds of quasiclassical variables 
that were sketched
in the Introduction. In our  family of coarse-grainings, the ones 
more useful for prediction are the more local
ones associated with an approximately conserved quantity. 
Our analysis of the harmonic chain 
is a step
towards a more realistic analysis of classicality in at least three
ways: (1) It considered a system which permits a continuum approximation
of the kind usually found in classical physics. (2) It employed 
a system-environment split which
follows from the coarse-grainings needed to realize that approximation
rather than being posited {\it ad hoc} in terms of fundamental
coordinates. (3) Different coarse-grainings were compared quantitatively
with respect to decoherence, noise and computational complexity. 

However, these positive features should not obscure how short this
analysis falls from the kind of treatment of classicality envisaged
by \cite{GH90a,GH93a,BH96a,PZ93,Ana99}. We considered only linear interactions,
not a realistic Hamiltonian. We did not compare all possible sets of
alternative coarse-grained histories, but only a four-parameter family
of them. We did not propose a unified quantitative measure for
classicality, but rather dealt separately with some of its attributes:
decoherence, resistance to noise, and computational simplicity. 
We did not start from the initial condition of the universe nor
exhibit the important role played by gravity in creating the conditions
for local equilibrium while ensuring the absence of global equilibrium.
Rather we assumed these properties in our initial condition.
Future analyses will do better. 


\acknowledgments
The authors are grateful for conversations with M. Gell-Mann over a long
period of time concerning the issues addressed in this paper,
as well as discussions with J.J. Halliwell. The 
work of both authors was supported in part by NSF grant PHY94-07194 
at the Institute for Theoretical Physics. That of JBH was 
also supported by NSF grant PHY95-07065, while TAB was also
supported by NSF grant PHY96-02084.

\appendix
\section{System-Environment Splits}

In this appendix we describe the circumstances under which---for a given
coarse-grained set of alternatives---the Hilbert space ${\cal H}$ of a
closed system can be written as a tensor product ${\cal H}^{s}
\otimes {\cal H}^{e}$ where ${\cal H}^{s}$ contains the
quantities followed by the coarse-graining and ${\cal H}^{e}$
contains the quantities that are ignored.
Such a tensor product factorization is called
a system-environment split. 

We begin by considering a Hilbert space ${\cal H}$ and a set of
alternatives at a single moment of time represented by an exhaustive set of
orthogonal projection operators $\{P_\alpha\}$, $\alpha=1,2,\cdots$
satisfying
\begin{equation}
P_\alpha P_\beta = \delta_{\alpha\beta}P_\beta\ , \quad
\sum_\alpha P_\alpha=I\ .
\label{aone}
\end{equation}
We seek to  write
\begin{equation}
{\cal H} = {\cal H}^s \otimes {\cal H}^e
\label{atwo}
\end{equation}
such that
\begin{equation}
P_\alpha = P^s_\alpha \otimes I^e\ , \quad \alpha=1,2,\cdots .
\label{athree}
\end{equation}
The decomposition (\ref{atwo}) is then a system-environment split.
If ${\cal H}^s$ were to contain {\it just} the quantities followed by the
coarse-graining we would naturally impose ${\rm dim}\,(P^s_\alpha)=1$, where 
${\rm dim}\,(P)$ is the dimension of the subspace projected on by $P$. However, other
notions of an environment can be useful in which $({\rm dim}\,P^s_\alpha)>1$, and
we shall consider the general case.

Only two simple mathematical facts are needed to analyze the above
question. First, the decomposition (\ref{athree}) requires the relation
between dimensions
\begin{equation}
 {\rm dim}\,\left(P_\alpha\right) =  {\rm dim}\,\left(P^s_\alpha\right)
{\rm dim}
\,\left({\cal H}^e\right) \ ,
\label{afour}
\end{equation}
and, as a special case
\begin{equation}
 {\rm dim}\,({\cal H}) =  {\rm dim}\,\left({\cal H}^s\right) 
{\rm dim}\,\left({\cal H}^e\right) \ .
\label{afive}
\end{equation}
Second, two separable Hilbert spaces of a given dimension are isomorphic
--- a consequence of the fact that they both have countable bases.

Realistic cases have ${\rm dim}\, ({\cal H})=\infty$, but we pause to note some
evident results from (\ref{afour}) when ${\rm dim}\, ({\cal H})$ is finite. Then
all the ${\rm dim}\, (P_\alpha)$ are finite. A system-environment split is not
always possible, only when the ${\rm dim}\, (P_\alpha)$
are all divisible by a common factor. In particular, if ${\rm dim}\,
(P^s_\alpha)$ is required to be unity, then a system-environment split is
possible only when all the $P$'s have the same dimension.

When ${\rm dim}\, ({\cal H})$ is infinite, there are a number  of subcases which
are convenient to treat separately. The most important of these is when
${\rm dim}\, (P_\alpha)=\infty$ for all $\alpha$. The requirements (\ref{aone})
imply that the $P_\alpha$ all commute and can be simultaneously
diagonalized.  Let $\{|i\rangle\}$, $i=1,2,\cdots$ be a basis in which they
are all diagonal. Then to each $P_\alpha$ there is a subset of these basis vectors
spanning the corresponding subspace. We write
\begin{equation}
P_\alpha = \sum_{i\in \alpha} |i\rangle\langle i|\ .
\label{asix}
\end{equation}
However, it is then a simple matter of relabeling to define an isomorphism
between the infinite dimensional Hilbert spaces
${\cal H}$ and ${\cal H}^s\otimes {\cal
H}^e$. We write
\begin{equation}
|i\rangle = |\alpha, A\rangle
\label{aseven}
\end{equation}
where $i$ ranges over the infinity of states in $P_\alpha$ and
$A=1,2,\cdots$ is another labeling of them. 
This relabeling defines the tensor product
${\cal H}^s \otimes {\cal H}^e$. Operators acting only on ${\cal H}^s$
have the form
\begin{mathletters}
\label{aeight}
\begin{equation}
\left\langle\alpha^\prime A^\prime|{\cal O}|\alpha A\right\rangle =
\left\langle\alpha^\prime|{\cal O}^s|\alpha\right\rangle\,
\delta_{AA^\prime}
\label{aeight a}
\end{equation}
while those on ${\cal H}^e$ have matrix elements
\begin{equation}
\left\langle\alpha^\prime A^\prime|{\cal O}|\alpha A\right\rangle =
\delta_{\alpha^\prime\alpha} \left\langle A^\prime|{\cal O}^e|A\right\rangle
\ .
\label{aeight b}
\end{equation}
\end{mathletters}
In particular, 
\begin{equation}
P_\alpha = P^s_\alpha \otimes I^e = (|\alpha\rangle\langle\alpha|)\otimes
I^e\ .
\label{anine}
\end{equation}
and ${\rm dim}\, (P^s_\alpha)=1$. We have constructed a system-environment split
defined by the coarse-graining $\{P_\alpha\}$.

If the condition ${\rm dim}\, (P^s_\alpha)=1$ is relaxed
it is possible to define
many other system-environment splits for this coarse-graining. One simply
relabels including more states in ${\cal H}^s$, {\it viz.}
\begin{equation}
|i\rangle = |a, A\rangle
\label{aten}
\end{equation}
such that the subspace $P^s_\alpha$ contains several different values
of $a$.

Many calculations use a system-environment split of this kind. For example,
in studies of Brownian motion the labels $a$ correspond to the coordinates
of the massive particle and $A$ to the coordinates of the particles of the
bath. The coordinates of the bath are ignored, but typically the
coordinates of the Brownian particle are followed only to some accuracy.
Thus, for a given choice of ranges, coordinates other than those in the bath
are ignored corresponding to ${\rm dim}\, (P^s_\alpha) > 1$. There is no unique
system-environment split.

The key to the above construction is that the relation (\ref{afour}) is
easily satisfied because ${\rm dim}\, (P_\alpha)$ and ${\rm dim}\, ({\cal H})$ are both
infinite. Finite dimensional members of the set of alternatives are
obstacles to a system-environment split. If ${\rm dim}\, (P_\alpha)$ is finite for
some $\alpha$, then (\ref{afour}) can only be satisfied if ${\rm dim}
\, (P^s_\alpha)$ and ${\rm dim}\, ({\cal H}^e)$ are both finite --- already
a restrictive condition. Furthermore, the dimensions of all the finite
dimensional $P_\alpha$ must be divisible by common factor --- 
${\rm dim}\, ({\cal
H}^e)$ --- and this is not always possible.  The
only case in which the condition ${\rm dim}\, (P^s_\alpha)=1$  could be enforced
is if all the finite dimensional $P_\alpha$ have the {\it same} dimension.
Clearly, a system-environment split is generally possible only 
when the dimensions of all of the $P_\alpha$ are infinite.

\section{Matrices for the Chain of Oscillators}

In this appendix we complete a calculation begun in Section IIIC.
This is the explicit demonstration that when the variables in the
action are changed according to (\ref{twonine}) and (\ref{twofourteen}),
the result yields the same equations of motion as are obtained by making
these changes in the fine-grained equations of motion directly. 
Specifically we check that when (\ref{threeeight}) is used to eliminate
the $q$'s from (\ref{threetwenty}) it yields (\ref{threethirteen}).  

Since the coarse-grained modes $A_L^{(d)}(t)$ are all decoupled from each other,
and interact with separate collections of high-frequency modes, we can
consider them one at a time, which somewhat simplifies our notation.
In this case, the matrix ${\bf S}$ reduces to a diagonal matrix
with elements making up a single vector with
$d$ components $S_a$ and ${\bf T}$ is a $d$ by $d-1$ matrix with components
$T_{ab}$, where $a$ ranges from $0$ to $d-1$ and $b$ ranges from $1$
to $d-1$.  Thus ${\bf M}_{SS}$ and ${\bf V}_{SS}$ become scalars,
and ${\bf M}_{ST}$, ${\bf M}_{TS}$, ${\bf V}_{ST}$ and ${\bf V}_{TS}$ are
vectors in the space of the $q$'s.
These matrices have the simple form
\begin{equation}
S_0 = {1\over c_0}, \ \ \ S_{a} = 0, \quad {\rm for}\ a>0
\label{bone}
\end{equation}
and
\begin{equation}
T_{0b} = -{c_b\over c_0}, \ \ \ T_{ab} = \delta_{ab}, \quad {\rm for} \  a>0
\label{btwo}
\end{equation}
where the numbers $c_a$ are the coefficients $c_{Lk}^{(d)}$ defined
by (\ref{twoeleven}) and (\ref{threesix}), forming the 
components of the vector ${\vec c}$
The matrix ${\bf \omega}$ is diagonal with components
$\omega_a$ and the mass is a constant $\mu$.

With these definitions, the matrices defined in
(\ref{threeeighteena}--\ref{threeeighteenb}) are
\begin{eqnarray}
M_{SS} &=& {\mu\over|c_0|^2}, \nonumber\\
({\bf M}_{TT})_{bb'} &=& \mu \delta_{bb'}
  + \mu {c_b c^*_{b'} \over |c_0|^2}, \nonumber\\
({\vec M}_{ST})_b &=& - \mu {c_b\over|c_0|^2} = ({\vec M}_{TS})_b, \nonumber\\
V_{SS} &=& \mu\omega_0^2, \nonumber\\
({\bf V}_{TT})_{bb'} &=& \mu \omega_b^2 \delta_{bb'}
  + \mu \omega_0^2 {c_b c^*_{b'} \over |c_0|^2}, \nonumber\\
({\vec V}_{ST})_b &=& - \mu\omega_0^2 {c_b\over|c_0|^2} = ({\vec V}_{TS})_b.
\label{bthree}
\end{eqnarray}
In these expressions when the matrices ${\bf M_{SS}}$, {\it etc.}, have
been reduced to scalars or vectors we have
changed the notation in what we hope is an obvious way. 
With (\ref{bthree}) we can now show that equation (\ref{threetwenty})
is identical to equation (\ref{threethirteen}) aside from a multiplicative
factor.  With a little algebra we find that
\begin{eqnarray}
(M_{SS} - {\vec M}_{ST}{\bf M}^{-1}_{TT}{\vec M}_{TS}) &=&
  \left({\mu\over{ |c_0|^2 + \sum_b |c_b|^2 }}\right), \nonumber\\
( V_{SS} - {\vec M}_{ST}{\bf M}^{-1}_{TT}{\vec V}_{TS}) &=&
  \left({\mu\over{ |c_0|^2 + \sum_b |c_b|^2 }}\right)\omega_0^2, \nonumber\\
- ({\vec V}_{ST} - {\vec M}_{ST}{\bf M}^{-1}_{TT}{\bf V}_{TT})_b &=&
  \left({\mu\over{|c_0|^2+\sum_{b'}|c_{b'}|^2}}\right)
  c_b(\omega_0^2 - \omega_b^2).
\label{bfour}
\end{eqnarray}
Since $\omega_0 \equiv \Omega_L$, plugging these back into
(\ref{threetwenty}) simply yields the equation
\begin{equation}
\left({\mu\over{|c_0|^2+\sum_b |c_b|^2}}\right)({\ddot A_L^{(d)}}(t)
  + \Omega_L^2 A_L^{(d)}(t) ) =
\left({\mu\over{|c_0|^2+\sum_b |c_b|^2}}\right)\Delta f(t),
\label{bfive}
\end{equation}
that is, simply equation (\ref{threethirteen}) multiplied by a constant
factor.  This factor is independent of the choice of coarse-grained
mode $L$ or the coarse-graining $d$:
\begin{equation}
\left( {\mu\over |c_0|^2 + \sum_b |c_b|^2} \right) =
  \left( {\mu\over |c_0|^2 + {\vec c}{\vec c}} \right) = N\mu\ .
\label{bsix}
\end{equation}

\section{Transforming the Equation of Motion}

In Section III we derived the classical equation of motion for
the coarse grained variables $A_L(t)$ utilizing two different
ways of eliminating the environmental coordinates $q_a$.  First,  
using the fact that the $q$'s were themselves fine-grained coordinates, 
we solved the classical equations of motion for the $q$'s and
substituted the solution into the fine-grained equations of motion.
Second, we derived the equations of motion from the Lagrangian
written in terms of the $A$'s and $q$'s and then solved the for
the $q$'s to eliminate them. In this appendix we complete the 
demonstration that the ostensibly different equations for the 
$A$'s that result are, in fact, equivalent.  

Suppose we have an equation of motion of the form (\ref{threeten})
for a single coarse-grained mode $A_L$,
which in this appendix we call ${\cal E}_L=0$. We wish to write it in 
a new form (\ref{threetwentythree}) which we call ${\cal E}'_L=0$
by a transformation of the form:
\begin{equation}
{\cal E}'_L(t,A_L(\tau)] = C\left( {\cal E}_L(t,A_L(\tau)]
  + \int_0^t dt'  G_L(t,t') {\cal E}_L(t',A_L(\tau)] \right)
  = \Delta f'_L(t),
\label{cone}
\end{equation}
with, of course, a transformed noise function
\begin{equation}
\Delta f'_L(t) = C\left( \Delta f_L(t)
  + \int_0^t dt' G_L(t,t') \Delta f_L(t') \right) \ . 
\label{ctwo}
\end{equation}
Here, $C$ is a positive constant and $G_L(t,t')$ is a particular Green's
function for the equation $ {\cal E}_L=0$, that is in
operator shorthand ${\cal E}_L G_L = I$.
In a similar shorthand we refer to the
transformation (\ref{cone}) as $C(I+G)$.  Clearly,
the solutions to the two equations (\ref{threeten}) and 
(\ref{cone}) will be the same only if $(I+G)$ is invertible.
This will be true if the equation
\begin{equation}
f_L(t) + \int_0^t G_L(t,t') f_L(t') dt' = 0
\label{cthree}
\end{equation}
can only be solved by $f_L(t)=0$.  {F}rom (\ref{cthree}) it is clear
that any solution $f_L(t)$ must have $f_L(0)=0$, and is a solution to
a second-order linear equation.

In the case of the chain of oscillators, the original form of the
equation is (\ref{threethirteen}) and the transformed equation is
(\ref{threetwentythree}).  {F}rom Appendix B we see that the positive
constant is
\begin{equation}
C = \left({\mu\over{|c_0|^2+{\vec c}{\vec c}}}\right) = N\mu,
\label{cfour}
\end{equation}
and from (\ref{threetwentythree}) and
(\ref{bfour}) the kernel $G_L(t,t')$ is
\begin{equation}
G_L(t,t') = {\mu\over|c_0|^2} {\vec c} (\Omega_L^2{\bf I} - {\bf\Omega}_Q^2)
  {\bf M}^{-1/2}_{TT} {\bf\Omega}^{-1} \sin[{\bf\Omega}(t-t')]
    {\bf M}^{-1/2}_{TT} {\vec c},
\label{cfive}
\end{equation}
where $\Omega_L^2$ is the constant
$\omega_0^2$ and ${\bf\Omega}_Q^2$ is
the diagonal matrix ${\bf\omega}$ restricted to the $q$'s, with
diagonal elements $\omega_b^2$.

{F}rom (\ref{cfive}) we know that $G_L(t,t)=0$ for all $t$, which
implies
\begin{equation}
{ d f_L\over dt}(t) = - \int_0^t {d G_L\over dt}(t,t') f_L(t') dt'
\label{csix}
\end{equation}
and hence $(df_L/dt)(0) = 0$.  Thus, the only solution to (\ref{cthree})
is $f_L(t)=0$ and therefore $(I+G)$ is indeed invertible.

Now we need to show that $C(I+G)\Delta f_L(t) = \Delta f'_L(t)$ for
our harmonic chain.  The two forms of the noise are given by
(\ref{threetwentyfoura}) and (\ref{threetwentyfourb}), respectively.
Let ${\bf\Omega}^2$ be the transformed frequency matrix defined by
(\ref{threetwentytwo}), with eigenvalues $\nu_k^2$ and orthonormal eigenvectors
${\vec v}_k$, and ${\bf\Omega}^2_Q$ be the diagonal frequency matrix
with eigenvalues $\omega^2_a$.  Then the requirement (\ref{ctwo})
can be written
\begin{eqnarray}
&& N\mu {\vec c}(\Omega_L^2{\bf I}-{\bf\Omega}_Q^2)
  \left(\cos({\bf\Omega}_Qt)\Delta{\vec q}(0) +
  {\bf\Omega_Q}^{-1} \sin({\bf\Omega}_Qt) \Delta{\dot{\vec q}}(0) \right) \nonumber\\
&+& {N\mu^2\over|c_0|^2} {\vec c}{\bf M}_{TT}^{-1/2}(\Omega_L^2{\bf I} -
{\bf\Omega}_Q^2)
  {\bf\Omega}^{-1}\int_0^t \sin({\bf\Omega}(t-t')){\bf M}_{TT}^{-1/2}{\vec c}
  \nonumber\\
&& \times {\vec c}(\Omega_L^2{\bf I}-{\bf\Omega}_Q^2)
  \left(\cos({\bf\Omega_Q}t)\Delta{\vec q}(0)
  + {\bf\Omega}_Q^{-1}\sin({\bf\Omega}_Qt)\Delta{\dot{\vec q}}(0)\right) dt' \nonumber\\
&=& N\mu {\vec c}(\Omega_L^2{\bf I}-{\bf\Omega}^2)
  \left(\cos({\bf\Omega} t){\bf M}_{TT}^{1/2} \Delta{\vec q}(0) +
  {\bf\Omega}^{-1}\sin({\bf\Omega} t){\bf M}_{TT}^{1/2}\Delta{\dot{\vec q}}(0) \right)\ .
\label{cseven}
\end{eqnarray}
By writing the vectors in terms of eigenvectors of ${\bf\Omega}^2$ we can
break the integral in (\ref{cseven}) into a sum over many integrals
having the forms
\begin{equation}
\int_0^t \sin(\nu(t-t'))\sin(\omega t') dt'
  = {1\over{\nu^2-\omega^2}}\left[ \nu\sin(\omega t)
  - \omega\sin(\nu t) \right]\ ,
\label{ceight}
\end{equation}
and
\begin{equation}
\int_0^t \sin(\nu(t-t'))\cos(\omega t') dt'
  = {\nu\over{\nu^2-\omega^2}}\left[ \cos(\omega t)
  - \cos(\nu t) \right]\ .
\label{cnine}
\end{equation}
Substituting these values into (\ref{cseven}),
the result we wish to show becomes
\begin{eqnarray}
&&\sum_a  c_a(\Omega_L^2-\omega_a^2)\left(\cos(\omega_at)\Delta q_a(0) +
  {1\over\omega_a} \sin(\omega_at) \Delta{\dot q}_a(0) \right) \nonumber\\
&+& {\mu\over|c_0|^2} \sum_{a,k} |{\vec c}{\bf M}_{TT}^{-1/2}{\vec v}_k|^2
  c_a {(\Omega_L^2 - \nu_k^2)(\Omega_L^2-\omega_a^2)\over(\nu_k^2-\omega_a^2)}
  \biggl[ \cos(\omega_at) \Delta q_a(0) \nonumber\\
&& - \cos(\nu_kt)\Delta q_a(0)
  + {1\over\omega_a}\sin(\omega_at) \Delta{\dot q}_a(0)
  - {1\over\nu_k}\sin(\nu_kt) \Delta{\dot q}_a(0) \biggr] \nonumber\\
&=& \sum_k ({\vec c}{\bf M}_{TT}^{-1/2}{\vec v}_k)
  (\Omega_L^2-\nu_k^2) \left(\cos(\nu_k t)
  ({\vec v}_k{\bf M}_{TT}^{1/2} \Delta{\vec q}(0)) +
  {1\over\nu_k}\sin(\nu_k t)
  ({\vec v}_k{\bf M}_{TT}^{1/2}\Delta{\dot{\vec q}}(0)) \right)\ .
\label{cten}
\end{eqnarray}
We wish this to hold at arbitrary times $t$, which implies that each
frequency must be equated separately.  This requires that the
following two conditions hold true:
\begin{mathletters}
\begin{equation}
{\bf I} +
  {\mu\over|c_0|^2} {\vec c}{\bf M}_{TT}^{-1/2}(\Omega_L^2{\bf I}-{\bf\Omega}^2)
  ({\bf\Omega}^2-{\bf\omega}_a^2{\bf I})^{-1}{\bf M}_{TT}^{-1/2}{\vec c} = 0\ ,
\label{celevena}
\end{equation}
\begin{equation}
({\vec v}_k{\bf M}_{TT}^{1/2}\Delta{\vec q}(0)) +
  {1\over |c_0|^2}({\vec v}_k{\bf M}_{TT}^{-1/2}{\vec c}) \sum_a c_a
  {(\Omega_L^2-\omega_a^2)\over(\nu_k^2-\omega_a^2)}\Delta q_a(0)\ .
\label{celevenb}
\end{equation}
\end{mathletters}

The first condition (\ref{celevena}) must hold for all $a$.
The last condition (\ref{celevenb}) must hold for arbitrary initial
vector $\Delta{\vec q}(0)$ and all $k$.

We can readily evaluate (\ref{celevena}) from the matrix definitions
(\ref{bthree}).  The inverse matrix is
\begin{equation}
({\bf\Omega}^2-\omega_a^2{\bf I})^{-1} = {\bf M}_{TT}^{1/2}
  ({\bf V}_{TT} - \omega_a^2 {\bf M}_{TT})^{-1} {\bf M}_{TT}^{1/2},
\label{ctwelve}
\end{equation}
and the matrix $({\bf V}_{TT} - \omega_a^2{\bf M}_{TT})$ has the
simple form ${\bf M}_{ij} = f_i\delta_{ij} + g_i g_j$ and hence
can be analytically inverted (even though one of the $f_i$ vanishes).
Carrying out this inverse and performing the sum, we see that the $a$
dependence of (\ref{celevena}) does indeed drop out, and the equation
is satisfied.

The second condition (\ref{celevenb}) looks even more difficult to evaluate,
since we have no explicit expressions for the eigenvalues $\nu_k^2$ and
eigenvectors ${\vec v}_k$.  However, these are not necessary.  Note that
if ${\vec v}_k$ is an eigenvector of ${\bf\Omega}^2 =
{\bf M}_{TT}^{-1/2} {\bf V}_{TT} {\bf M}_{TT}^{-1/2}$ with eigenvalue
$\nu_k^2$, then ${\bf M}_{TT}^{1/2}{\vec v}_k$ is an eigenvector
of ${\bf V}_{TT}{\bf M}_{TT}^{-1}$ with the same eigenvalue.  Using
the definitions (\ref{bthree}), this implies that
\begin{equation}
({\bf M}_{TT}^{1/2}{\vec v}_k)_a =
  {c_a(\Omega_L^2-\omega^2_a)\over(\nu_k^2-\omega^2_a)(|c_0|^2
  + {\vec c}{\vec c})}({\vec c}{\bf M}_{TT}^{1/2}{\vec v}_k)\ .
\label{cthirteen}
\end{equation}
Since ${\vec c}$ is an eigenvector of ${\bf M}_{TT}$, we can readily show
\begin{equation}
({\vec v}_k{\bf M}_{TT}^{-1/2}{\vec c}) = {1\over\mu}
  {|c_0|^2\over |c_0|^2+{\vec c}{\vec c}}
  ({\vec v}_k{\bf M}_{TT}^{1/2}{\vec c})\ .
\label{cfourteen}
\end{equation}
By combining (\ref{cthirteen}) and (\ref{cfourteen}), the condition
(\ref{celevenb}) is immediately proven.  Hence, the transformed
classical noise is the same as the noise derived from a straightforward
Lagrangian calculation.

This result has immediate implications for the quantum probabilities
(\ref{sixtwentyfive}).  The integrand in the exponent of that expression
is proportional to 
\begin{equation}
{\cal E}'_L(t,A_L(\tau)] {\rm M}[\Delta f'(t)\Delta f'(t')]^{\rm inv}
  {\cal E}'_L(t',A_L(\tau)]
\label{cfifteen}
\end{equation}
summed over all $L$.  This is all in terms of the equation of motion form
(\ref{threetwentythree}) and noise correlation function
(\ref{threetwentyseven}).  But since ${\cal E}'_L = C(I+G){\cal E}_L$
and ${\rm M}[\Delta f'(t)\Delta f'(t')]^{\rm inv} =
C^{-1}(I+G)^{-1} {\rm M}[\Delta f(t)\Delta f(t')]^{\rm inv} C^{-1}(I+G)^{-1}$,
we see that the transformation $C(I+G)$ cancels out, and
\begin{equation}
{\cal E}'_L(t,A_L(\tau)] {\rm M}[\Delta f'(t)\Delta f'(t')]^{\rm inv}
  {\cal E}'_L(t',A_L(\tau)] =
{\cal E}_L(t,A_L(\tau)] {\rm M}[\Delta f(t)\Delta f(t')]^{\rm inv}
  {\cal E}_L(t',A_L(\tau)]\ ,
\label{csixteen}
\end{equation}
{\it i.e.}, even in the quantum case one can use the simpler form of the
classical equation of motion and noise.


\end{document}